\documentclass[aps,journal=jcp,twocolumn,noeprint,superscriptaddress,amsmath,amssymb,showpacs,nolongbibliography]{revtex4-2}
\usepackage{graphicx}
\usepackage{amsmath}
\usepackage{graphics}

\usepackage{amsmath}
\usepackage{amsfonts}
\usepackage{amssymb}
\usepackage{xcolor}
\graphicspath{{figs/}}

\usepackage[caption=false]{subfig}

\usepackage[bookmarks=false]{hyperref}
\hypersetup{colorlinks=true, citecolor=blue, urlcolor=blue, linkcolor=blue}

\newcommand{\pd}{\phantom{\dagger}}
\newcommand{\figref}[1]{\ref{#1}}

\usepackage[normalem]{ulem}

\begin{document}

\title{Improving the Stability of the Hierarchical Equations of Motion for Open Quantum Systems with Strong Coupling to Structured Bosonic Baths}

\author{Salvatore Gatto $^{\dag}$}
\affiliation{Institute of Physics, University of Freiburg, Hermann-Herder-Str. 3, D-79104 Freiburg, Germany}
\email{Email: salvatore.gatto@physik.uni-freiburg.de}
\thanks{{\dag} S.G and S.L.R contributed equally to this work.}
\author{Samuel \ L.\ Rudge$^{\dag}$}
\affiliation{Institute of Physics, University of Freiburg, Hermann-Herder-Str. 3, D-79104 Freiburg, Germany}
\author{Bokang Hou}
\affiliation{Division of Chemistry and Chemical Engineering, California Institute of Technology, Pasadena, California 91125, United States of America}
\author{Eran Rabani}
\affiliation{Fritz Haber Center for Molecular Dynamics,	The Institute of Chemistry and The Institute of Applied Physics, The Hebrew University of Jerusalem, Jerusalem 91904, Israel}
\affiliation{Department of Chemistry, University of California, Berkeley, California 94720, United States of America}
\author{Michael Thoss}
\affiliation{Institute of Physics, University of Freiburg, Hermann-Herder-Str. 3, D-79104 Freiburg, Germany}

\begin{abstract}
	\noindent The hierarchical equations of motion (HEOM) approach is one of the most powerful methods available to simulate the dynamics of open quantum systems. Although the HEOM approach has been shown to be stable and efficient in many parameter regimes and physical scenarios, it is also well known that the unavoidable finite truncation of the hierarchy can lead to fundamental numerical instabilities for bosonic environments. In this work, we apply recently developed techniques to analyze these issues. We use an auxiliary Fock-space picture of the hierarchy to first identify the unbalanced hierarchy-raising terms responsible for the strong transient amplification associated with the non-normal algebraic structure of the HEOM generator. We then apply a non-unitary similarity transformation that converts these terms into balanced combinations of hierarchy-raising and hierarchy-lowering terms, thereby suppressing their numerical impact. We first demonstrate the efficacy of this approach via a spin-boson model with a simple Brownian oscillator spectral density and subsequently extend the analysis to a structured spectral density, showing for both cases that the transformed HEOM approach allows simulations of much larger system-bath coupling than would otherwise be possible.
\end{abstract}

\maketitle

\newpage

\section{Introduction}

\noindent Any quantum system interacting with its surroundings can undergo energy and particle exchange with external environments, leading to irreversible processes such as dissipation and decoherence.
Consequently,  open quantum system theory is central to a broad range of problems, from coherence and control of qubits \cite{vanderSar2012_decoherence-protected,Naeij2025_open,Onizhuk2025_colloquium}, to electron, proton, and energy transfer in the condensed phase	\cite{Ke2022_nonequilibrium,Shi2009_electron,Slocombe2022_an_open,Liu2021_understanding,Zhang2020_proton,Ishizaki2009_energy}, as well as vibronic dynamics in molecular and solid-state materials \cite{Hou2025_unraveling,Bai2024_hierarchical,Schinabeck2016_hierarchical,Erpenbeck2019_hierarchical,Rudge2024_nonadiabatic,Tamura2013_charge,Chan2024_excitonphonon,Varvelo2023_formally}. Simultaneously, many of these problems occur in challenging non-perturbative regimes, with strong system-bath coupling, low temperature, or highly structured spectral densities. As a result, the development and implementation of sophisticated theoretical methods capable of treating such complicated systems is a critical task.

Due to this high interest, many powerful approaches have been developed that can rigorously treat the non-Markovian and non-perturbative dynamics of open quantum systems  \cite{Wang2003_multilayer,Wang2009_numerically,Wang2015_multilayer,Prior2010_efficient,Chin2010_exact,Makri1992_improved,Ilk1994_real,Stockburger2002_exact,Suess2014_hierarchy,Strathearn2018_efficient,Muhlbacher2008_realtime,Cohen2011_memory}. Of these various methods, the HEOM approach \cite{Tanimura1989,Tanimura1990_nonperturbative,Ishizaki2005_quantum,Xu2007_dynamics,Shi2009_efficient,Shi2018_efficient,Batge2021_nonequilibrium,Tanimura2020} has emerged as one of the most versatile and successful techniques for modeling the dynamics of systems linearly coupled to noninteracting harmonic environments, for which it provides a numerically exact representation of the
Feynman-Vernon influence functional \cite{Takahashi2024_tensor-train,Ke2022_hierarchical,Jin2007_dynamics,Gatto2025_quantum}.



However, despite these numerous successes, there is growing evidence of intrinsic numerical instabilities within the bosonic hierarchy \cite{Reichman2019_removing,Yan2020_a_new,Li2022_a_low-temperature,Krug2023_on_stability,Xu2023_about}.
 Specifically, a key element of the HEOM approach is the non-perturbative expansion of the Feynman-Vernon influence functional in terms of a hierarchy of auxiliary density operators (ADOs). For bosonic baths, the resulting hierarchy is technically infinite, such that practical simulations require a truncation of the bosonic hierarchy at some finite tier $n_{\text{max}}$, which does not necessarily preserve the intrinsic stability of the exact Gaussian influence functional. 


Despite this connection to a finite truncation tier, the instabilities cannot always be resolved by increasing $n_{\text{max}}$, as has been shown for a variety of simple spectral densities with strong system-bath couplings \cite{Reichman2019_removing,Yan2020_a_new,Li2022_a_low-temperature,Krug2023_on_stability}. Furthermore, this problem is exacerbated by the advent of sophisticated pole decomposition schemes \cite{Hu2011_pade,Chen2022_prony,Nakatsukasa2018_the_AAA,	Roy1989_ESPRIT,Takahashi2024_high,Mack2024_nonadiabatic} capable of efficiently representing structured spectral densities, which, as we show in this work, can induce numerical instabilities even at moderate coupling strength. Given the obvious interest in modeling realistic systems, removing such instabilities remains an important task of the field. 

So far, a variety of methods have been proposed to minimize instabilities within the bosonic HEOM approach. These include convergence attempts with large truncation tier \cite{Shi2018_efficient}, projection methods via Prony filtration of unstable eigenvalues in the propagator \cite{Reichman2019_removing}, and equivalent representations of the HEOM \cite{Ikeda2020_generalization,Yan2020_a_new,Li2022_a_low-temperature,Ding2012_optimizing,Duan2017_zero-temperature,Liu2014_reduced}. In Ref.~\cite{Ikeda2020_generalization}, for example, a generalized form of the HEOM that includes non-exponential components in the bath-correlation expansion was introduced, resolving instabilities for the critically damped Brownian oscillator spectral density. Meanwhile, in Ref.~\cite{Yan2020_a_new}, a similar equivalence between the HEOM and the mixed quantum-classical Liouville equation \cite{Liu2014_reduced} was also applied to discrete harmonic oscillator modes, with the approach further extended in Ref.~\cite{Li2022_a_low-temperature} to Brownian oscillator spectral densities and yielding a stable HEOM in the form of a quantum Fokker-Planck equation. Related changes of representation in extended auxiliary spaces, including Bogoliubov transformations between different Markovian embeddings, have also been discussed in Refs.~\cite{Xu2026_markovian,Xu2026_simulating}. Here, we focus specifically on the effect of such representation changes on non-normal amplification and numerical stability under finite hierarchy truncation.


Building on the idea of more stable representations of HEOM, in this work we extend the methodology of Ref.~\cite{Li2022_a_low-temperature} to arbitrary spectral densities. In contrast to previous work, we do not transform to a phase-space picture, but instead apply a non-unitary similarity transformation directly in the auxiliary Fock space, followed by a Bogoliubov-like rotation of the transformed hierarchy. This procedure converts the creation-like couplings responsible for large pseudospectral amplification into balanced combinations of hierarchy-raising and hierarchy-lowering terms, thereby substantially suppressing the  instabilities. We demonstrate this approach first for a simple Brownian oscillator spectral density in a spin-boson model, before simulating a spin-boson model with a highly structured spectral function, which has recently been employed to investigate exciton trap dynamics in quantum dots \cite{Hou2025_unraveling}. Our results show that the stabilized hierarchy can be used to investigate moderate to strong system-bath coupling even in the presence of complex, significantly non-Markovian spectral structure, broadening the applicability of HEOM to experimentally relevant systems.

The paper is structured as follows. We first introduce the spin-boson model and associated spectral functions in Sec.~\ref{sec: Model}. Next, in Sec.~\ref{sec: Hierarchichal Equations of Motion}, we briefly introduce the HEOM approach for bosonic environments, before discussing the fundamental stability of such methods and potential improvements to it in Sec.~\ref{subsec: Stability Analysis} and Sec.~\ref{subsec: Stability Improvement}, respectively. Finally, we demonstrate the stability improvement in Sec.~\ref{sec: Results}, first for a simple Brownian oscillator in Sec.~\ref{subsec: Brownian Oscillator} and then for a more realistic, structured spectral density in Sec.~\ref{subsec: Structured Spectral Density}, before concluding in Sec.~\ref{sec: Conclusion}. Note that throughout this work, we use units where $e = \hbar = 1$.

\section{Model} \label{sec: Model}

\noindent In this section, we introduce the general model considered in this work as well as the specific spectral functions used to demonstrate the more stable HEOM approach. Additionally, we provide some numerical details on efficient representations of bath-correlation functions, which is necessary to treat the structured spectral densities forming the main motivation of this work. 

We consider a general model of an open quantum system coupled to a bosonic bath, 
\begin{align}
H = \: & H_{\text{S}} + H_{\text{B}} + H_{\text{SB}},
\end{align}
where $H_{\text{S}}$ is the Hamiltonian of the quantum system, $H_{\text{B}}$ is the Hamiltonian of the bosonic bath, and $H_{\text{SB}}$ is the interaction between the two. In this work, we investigate a spin-boson model, 
\begin{align}
    H_{\text{S}} = \: & \varepsilon \sigma_{z} + \Delta \sigma_{x}, \label{eq: SP Ham}
\end{align}
where $\sigma_{i}$ refer to the Pauli matrices. Eq.\eqref{eq: SP Ham} describes a system with diabatic states $|n\rangle \in \{|0\rangle,|1\rangle\}$, defined as the eigenstates of $\sigma_{z}$, with corresponding diabatic energies $-\varepsilon$ and $+\varepsilon$, respectively, and diabatic coupling $\Delta$. 

This is coupled to a noninteracting bath of harmonic vibrational modes, 
  \begin{align}
    H_{\text{B}} = \sum_k \omega^{\pd}_k b_k^\dagger b^{\pd}_k.
\end{align} 
Here, $\omega^{\pd}_{k}$ refers to the frequency of mode $k$, while $b^{\dag}_{k}$ and $b^{\pd}_{k}$ are dimensionless operators that create and annihilate a phonon with energy $\omega_{k}$ in mode $k$, respectively. Finally, the interaction Hamiltonian is given by
\begin{align}
    H_{\text{SB}} = \: & O_{S} \otimes O_{B},
\end{align} 
where $O_{S}$ is an operator in the system Hilbert space determining the form of the system-bath coupling, while
\begin{align}
    O_{B} = \: & \sum_{k} \lambda_{k}\left(b^{\pd}_{k} + b^{\dag}_{k}\right)
	\end{align}
is the corresponding bath operator, with the coupling strength to mode $k$ determined by $\lambda_{k}$. For the spin-boson calculations considered in this work, we set $O_S=\sigma_z$, although the theory is applicable to a general system coupling operator $O_S$. Note that the theory is also applicable to multiple independent Gaussian
coupling channels, $H_{\mathrm{SB}} = \sum_\alpha O_{S,\alpha} \otimes O_{B,\alpha}$, such as for the case of multiple baths.

The spin-boson model is one of the most widely studied models in open quantum systems theory, as it represents one of the simplest nontrivial quantum systems coupled to a heat bath \cite{Leggett1987_dynamics,Weiss1999,Grifoni1998_driven,Bulla2003_numerical,ThossWang2008,Prior2010_efficient}.
This allows it to be used as a reliable benchmark for the accuracy, stability, and efficacy of quantum dynamical methods, which makes it highly suitable for this investigation. However, despite its apparent simplicity, it has been remarkably successful in modeling a wide variety of physical scenarios. Furthermore, although it forms a useful model for this work to benchmark improvements to the stability of the HEOM approach, we note that these improvements and HEOM in general can be applied to a much larger class of open quantum systems with multiple states and more exotic system-bath couplings. The only requirement is that the environmental influence is Gaussian and that the  system-bath coupling is linear in the bath operators.

The state of the total system is completely described by the total density matrix, $\rho(t)$. In the following, it is assumed that the system and bath are initially uncoupled, such that $\rho(0)$ factorizes, 
\begin{align}
    \rho(0) = \: & \rho_{\text{S}}(0) \otimes \rho_{\text{B}}(0), \label{eq: initial condition}
\end{align}
and the bath is assumed to initially be prepared in local thermal equilibrium at temperature $T$:
\begin{align}
    \rho_{\text{B}}(0) = \frac{e^{-\beta H_{\text{B}}}}{\text{Tr}_{\text{B}}\left\{e^{-\beta H_{\text{B}}}\right\}},
\end{align}
with the inverse temperature given by $\beta = 1 / k_{B}T$. For times $t > 0$, the influence of the bath on the evolution of the open quantum system is obtained by tracing out the bath degrees of freedom, and is completely characterized by the two-time bath-correlation function,
\begin{align}
    C(t) = \: & \text{Tr}_{\text{B}} \left\{O_{B}(t) O_{B}(0) \rho_{\text{B}}(0)\right\}.
\end{align}
Here, $O_{B}(t) = e^{i H_{\text{B}}t}O_{B}e^{-i H_{\text{B}}t}$ denotes the bath operator in the interaction picture. Equivalently, the bath-correlation function can also be expressed as  
\begin{align}
	C(t) = \frac{1}{\pi}\int_0^\infty d\omega\,J(\omega) \left[\coth\left(\frac{\beta \omega}{2}\right)\cos\left(\omega t\right) -i \sin\left(\omega t\right)\right].
\end{align} 
which has been written in terms of the system-bath spectral density,
\begin{align}
	J(\omega) = \pi \sum_k |\lambda_k|^2	\delta(\omega-\omega_k), \qquad \omega>0.
\end{align} 
In the following, it will be useful to write $C(t)$ as a two-sided integral, which can be achieved by extending the positive-frequency spectral density to negative frequencies through the odd continuation $J(-\omega)=-J(\omega)$:
\begin{align}
    C(t) = \frac{1}{2\pi} \int^{\infty}_{-\infty} d\omega \: J(\omega) \left[\coth\left(\frac{\beta\omega}{2}\right) + 1\right] e^{-i \omega t}.
\end{align}

\begin{figure}
    \vspace{-10mm}
    \begin{center}
       \includegraphics[width=\columnwidth]{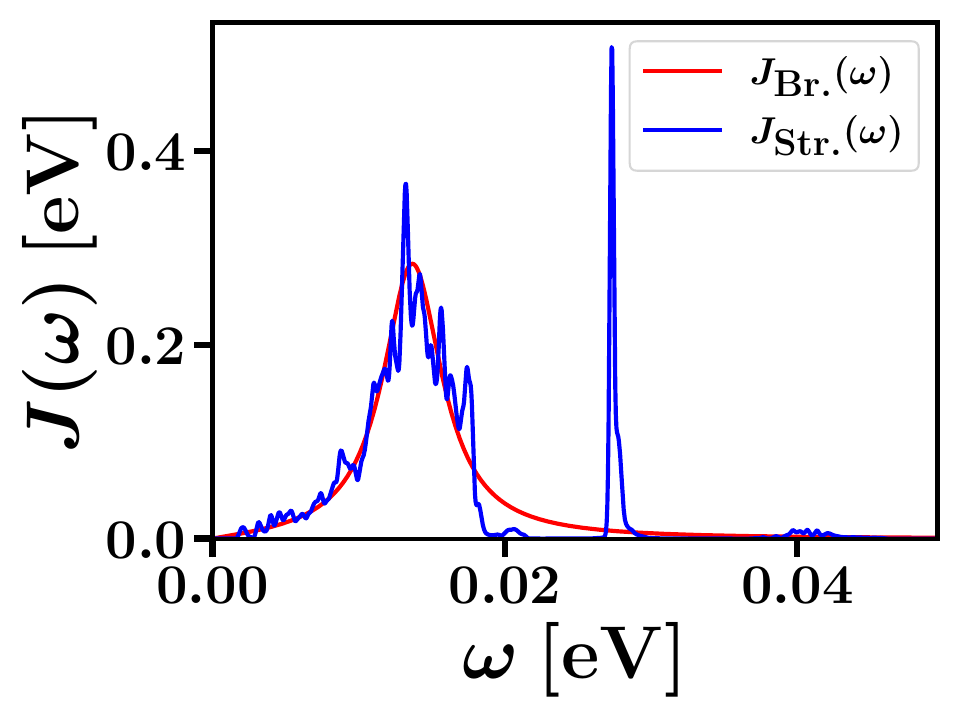}
       \caption{Brownian (red) and structured (blue) spectral density functions used in this work. Parameters of the Brownian oscillator are $\gamma = 4.9\text{ meV}$ and $\Omega = 13.89\text{ meV}$, and the reorganization energy of both has been set to $\Lambda = 50\text{ meV}$.}
       \label{fig: spectral functions}
    \end{center}
\end{figure}

Consequently, the spectral density $J(\omega)$ completely characterizes the bath and its influence on the system. In this work, two different spectral densities are investigated, representing two distinct physical scenarios. First, the system will be modeled as a damped Brownian oscillator, for which the spectral density is given by
\begin{align}
    J_{\text{Br.}}(\omega) = \: & 2\Lambda \frac{\gamma \Omega^2 \omega}{(\omega^{2} - \Omega^{2})^{2} + (\gamma \omega)^{2}},
\end{align} 
with natural frequency, $\Omega$, damping coefficient $\gamma$, and reorganization energy 
\begin{align}
    \Lambda = \: & \frac{1}{\pi} \int^{\infty}_{0} d\omega \: \frac{J(\omega)}{\omega}, \label{eq: renormalization energy definition}
\end{align}
which is a measure of the system-bath coupling strength. Additionally, we will also investigate a more structured spectral density, $J_{\text{Str.}}(\omega)$, which was originally introduced in Ref.~\cite{Hou2025_unraveling} to model nonradiative loss pathways in colloidal quantum dots. In this physical setting, the broad low-frequency band is associated with surface acoustic modes, whereas the sharp higher-frequency peak is associated with localized optical modes. Although the exact physical origin of this spectral function is not critical to the methodology of this paper, it is important to note that it represents typical structures found in more realistic spectral functions, with many peaks of varying width that cannot be simply described with a small number of Brownian oscillators. These two spectral functions are shown in Fig.~\figref{fig: spectral functions} for example parameters, where the reorganization energy of both spectral densities has been chosen as $\Lambda = 50\text{ meV}$. Furthermore, the natural frequency and damping coefficient of the Brownian oscillator have been chosen to mimic the first group of peaks in $J_{\text{Str.}}(\omega)$.

To end this section, we discuss how the bath-correlation function connects to the HEOM approaches introduced in Sec.~\ref{sec: Hierarchichal Equations of Motion}. Critically, methods such as HEOM \cite{Tanimura2020} include the non-Markovian, time-nonlocal influence of the bath on the system with Markovian embedding techniques \cite{Tanimura1989,Xu2022_taming}, in which the bath-correlation function is represented as a sum over exponential functions,
\begin{align}
    C(t) \approx \: & \tilde{C}(t) = \sum_{\ell}^{N^{\text{max}}_{\ell}} \eta_{\ell} e^{-\gamma_{\ell}t}. \label{eq: exponential decomposition of BCF}
\end{align}
Evidently, a key factor in the efficiency of such methods is obtaining an accurate representation of $C(t)$ for as few exponential terms as possible. Indeed, $N^{\text{max}}_{\ell}$ is generally treated as a convergence parameter within the HEOM approach. As a result, there has recently been great interest in developing efficient decomposition schemes for bath-correlation functions \cite{Hu2011_pade,Chen2022_prony,Nakatsukasa2018_the_AAA,
	Roy1989_ESPRIT,Takahashi2024_high,Mack2024_nonadiabatic}, especially in connection to structured spectral densities, since the number of terms needed to accurately represent $C(t)$ is determined by the temperature and the complexity of the associated $J(\omega)$. 

In this work, we will use two schemes. The first is the approach introduced in Ref.\cite{Xu2022_taming}, in which the temperature-broadened spectral function $S_{\beta}(\omega)$, 
\begin{align}
    S_{\beta}(\omega) = \: & J(\omega) \left[\coth\left(\frac{\beta\omega}{2}\right) + 1\right],
\end{align}
is decomposed directly in frequency space into a sum over rational functions via the $AAA$ algorithm \cite{Nakatsukasa2018_the_AAA}. The exponential form in Eq.\eqref{eq: exponential decomposition of BCF} is then obtained via residue theory. We found that this approach works well for the Brownian oscillator spectral density, with convergence generally reached at $N^{\text{max}}_{\ell} = 3$. In contrast, a two-stage procedure is required for $J_{\text{Str.}}(\omega)$, as outlined in Ref.\cite{Takahashi2024_high}. First, a highly accurate representation of $C(t)$ is obtained via the $AAA$ approach with a large number of terms. This is then used to construct the exact bath-correlation function $C(t)$, which is then decomposed up to a relevant time $t_{\text{max}}$ via an ESPRIT algorithm, which decomposes complex-valued time signals \cite{Takahashi2024_high}:
\begin{align}
    S_{\beta}(\omega) \overset{AAA}{\longrightarrow} C(t) \overset{\text{ESPRIT}}{\longrightarrow} \tilde{C}(t) .
\end{align}
The convergence of this bath-correlation function is discussed in further detail in Sec.~\ref{subsec: Structured Spectral Density}.


\section{Hierarchical Equations of Motion} \label{sec: Hierarchichal Equations of Motion}

\noindent In this section, we first provide an overview of the HEOM approach and the general equations of motion that serve as a starting point for the stability analysis discussed in Sec.~\ref{subsec: Stability Analysis}. We then present in Sec.~\ref{subsec: Stability Improvement} the stabilization approach introduced in Refs.~\cite{Yan2020_a_new,Li2022_a_low-temperature}, based on a non-unitary similarity transformation followed by a Bogoliubov transformation. Finally, in Sec.~\ref{subsec: Stabilized Algorithm}, we apply this approach to derive an explicit algorithmic formulation of the stabilized HEOM.

Within the HEOM approach, the object of interest is the reduced density matrix of the quantum system, which is obtained by tracing out the bath degrees of freedom from the total density matrix, $\rho_{\text{S}}(t) = \text{Tr}_{\text{B}}\left\{\rho(t)\right\}$. The key insight of HEOM is that the exponential form of the bath-correlation function in Eq.\eqref{eq: exponential decomposition of BCF} allows a Markovian embedding of the non-Markovian bath influence on the system degrees of freedom, resulting in a coupling of $\rho_{\text{S}}(t)$ to a series of auxiliary density operators (ADOs), $\rho^{(n)}_{\mathbf{j}}$, which also incorporate higher-order bath interaction effects. Here, $n$ is the so-called ADO tier, representing the order of system-bath interaction it describes, while $\mathbf{j} = (j_{1},\dots,j_{k},\dots,j_{n})$ with $j_{k} = (\ell_{k},s_{k})$ is a super-index detailing the specific bath-correlation poles used to construct this ADO and whether system operators act on it from the left, $s_{k} = 0$, or right, $s_{k} = 1$. Explicitly, the ADOs are related by a recursive relation, which has the following form for the spin-boson model,
\begin{align}
	\rho^{(n+1)}_{\mathbf{j}^{+}} = \: & \mathcal{B}^{\pd}_{j} \rho^{(n)}_{\mathbf{j}} \\
	= \: & -i \int^{t}_{0} d\tau \left(\eta^{\pd}_{\ell} e^{-\gamma^{\pd}_{\ell}(t-\tau)}O_{S} \rho^{(n)}_{\mathbf{j}}(\tau)\delta_{s 0} - \right. \nonumber \\
	& \qquad \qquad \:\:\:\:\: \left. \eta^{*}_{\ell} e^{-\gamma^{*}_{\ell}(t-\tau)}\rho^{(n)}_{\mathbf{j}}(\tau)O_{S}\delta_{s 1}\right),
\end{align} 
where $\mathbf{j}^{+} = (j_{1},\dots,j_{n},j)$ and the zeroth-tier ADO is simply the reduced system density matrix: $\rho^{(0)} = \rho_{\text{S}}$. 

Formally, this Markovian embedding is achieved by expressing the effect of the bath as an effective action in the Feynman-Vernon influence functional, which, after exploiting the self-similarity of the exponential representation of the bath-correlation function in Eq.\eqref{eq: exponential decomposition of BCF}, can in turn be written as a hierarchy of coupled equations of motion:
\begin{align}
    \frac{\partial}{\partial t}\rho^{(n)}_{\mathbf{j}} = \: & - \left(i\mathcal{L}_{\text{S}} + \sum_{k = 1}^{n} \gamma_{j_{k}} \right)\rho^{(n)}_{\mathbf{j}} \nonumber \\
    & - i \sum_{k = 1}^{n} \mathcal{C}_{j_{k}} \rho^{(n-1)}_{\mathbf{j}^{-}_{k}}  - i \sum_{j} \mathcal{A}_{j} \rho^{(n+1)}_{\mathbf{j}^{+}}. \label{eq: standard HEOM}
\end{align}
Here, $\mathcal{L}_{\text{S}} = [H_{\text{S}},\dots]$ is the generator of the system dynamics. Furthermore, a new super-index has been introduced, $\mathbf{j}^{-}_{k} = (j_{1},\dots,j_{k-1},j_{k+1}\dots,j_{n})$, which denotes the removal of a multi-index $j_{k}$, while the sum over $j$ in the final term represents the sum over all possible   $j = (\ell,s)$. The $j$-index frequencies are related to the original pole decomposition by $\gamma^{\pd}_{j=(\ell,0)} = \gamma_{\ell}$ and $\gamma_{j=(\ell,1)} = \gamma^{*}_{\ell}$. 

Finally, the final two terms couple ADOs between tiers, which are moderated by the coupling-up and coupling-down superoperators, respectively,
  \begin{align}
    \mathcal{A}_{j}\rho^{(n)}_{\mathbf{j}}  = \: & \left[O_{S},\rho^{(n)}_{\mathbf{j}}\right] \\
    \mathcal{C}_{j}\rho^{(n)}_{\mathbf{j}} = \: & \eta_{\ell}^{\pd}O_{S}\rho^{(n)}_{\mathbf{j}}\delta_{s 0} - \eta_{\ell}^{*}\rho^{(n)}_{\mathbf{j}}O_{S}\delta_{s 1}.
\end{align} 

\subsection{Stability Analysis} \label{subsec: Stability Analysis}

The goal of this and the following subsections is to find an equivalent representation of Eq.~\eqref{eq: standard HEOM} from which the terms responsible for finite-truncation instabilities can be identified. Our analysis follows the representation-based strategy adopted in previous studies of HEOM stability, including Refs.~\cite{Ikeda2020_generalization,Li2022_a_low-temperature}. Rather than reformulating the dynamics in a phase-space picture, we analyze the HEOM generator entirely within the auxiliary Fock space of the hierarchy, without assuming a particular form of the spectral function. This exposes the competition between damping terms, which suppress large auxiliary occupations, and creation-like hierarchy-raising terms, which transport weight towards higher tiers. The resulting operator structure provides a direct route to a non-unitary similarity transformation and a subsequent Bogoliubov-like reorganization designed to reduce transient amplification in finite truncations.

To expose the structural origin of numerical instabilities in the truncated HEOM approach, we first separate the damping terms, which are proportional to the occupation numbers of the auxiliary Fock-space modes and governed by the real parts of the bath-correlation exponents, from the oscillatory mixing between cosine and sine auxiliary sectors, governed by their imaginary parts.
 This is achieved by splitting the frequencies of the bath-correlation function into real and imaginary components, $\gamma_{\ell} = \gamma_{\ell,r} + i \gamma_{\ell,i}$, yielding 
  \begin{align}
\mathcal{B}_{j} = \: & \underbrace{\mathcal{B}_{(g=\cos,\ell,s)}}_{=\mathcal{B}_{j|}} + \underbrace{\mathcal{B}_{(g=\sin,\ell,s)}}_{=\mathcal{B}_{|j}},
\end{align} 
where $\mathcal{B}_{j_{\text{c}}|}$ and $\mathcal{B}_{|j_{\text{s}}}$ are in turn used to split the single multi-index of each ADO into a double-side index, $\rho^{(n)}_{\mathbf{j}_{\text{c}}|\mathbf{j}_{\text{s}}}$, defined via the recursive relations 
  \begin{align}
    \mathcal{B}^{\pd}_{j|}\rho^{(n)}_{\mathbf{j}_{\text{c}}|\mathbf{j}_{\text{s}}} = \: & -i \eta^{\pd}_{\ell} \int^{t}_{0} d\tau e^{-\gamma_{\ell,r}(t-\tau)}\cos\left(\gamma_{\ell,i}(t-\tau)\right) \times \nonumber \\
    & \qquad\qquad \left(O_{S} \rho^{(n)}_{\mathbf{j}_{\text{c}}|\mathbf{j}_{\text{s}}}(\tau)\delta_{s 0} - \rho^{(n)}_{\mathbf{j}_{\text{c}}|\mathbf{j}_{\text{s}}}(\tau)O_{S}\delta_{s 1}\right) \\
    \mathcal{B}^{\pd}_{|j}\rho^{(n)}_{\mathbf{j}_{\text{c}}|\mathbf{j}_{\text{s}}} = \: & -\eta^{*}_{\ell} \int^{t}_{0} d\tau e^{-\gamma_{\ell,r}(t-\tau)}\sin\left(\gamma_{\ell,i}(t-\tau)\right) \times \nonumber \\
    & \qquad\qquad \left(O_{S} \rho^{(n)}_{\mathbf{j}_{\text{c}}|\mathbf{j}_{\text{s}}}(\tau)\delta_{s 0} - \rho^{(n)}_{\mathbf{j}_{\text{c}}|\mathbf{j}_{\text{s}}}(\tau)O_{S}\delta_{s 1}\right).
\end{align} 

Next, we demonstrate that the stability properties of the HEOM emerge naturally when the discrete hierarchy is reformulated by introducing a Fock representation of the ADOs,
\begin{align}
    \rho^{(n)}_{\mathbf{j}_{\text{c}}|\mathbf{j}_{\text{s}}} \rightarrow \rho_{\substack{\mathbf{m}|\mathbf{n}\\\mathbf{o}|\mathbf{p}}},
\end{align}
where $\mathbf{m} = (m_{0},\dots,m_{N^{\text{max}}_{\ell}})$ and $\mathbf{n} = (n_{0},\dots,n_{N^{\text{max}}_{\ell}})$ denote vectors of occupation numbers counting the occurrences of $\mathcal{B}_{(\cos,\ell,0)}$ and $\mathcal{B}_{(\sin,\ell,0)}$, respectively, while $\mathbf{o}$ and $\mathbf{p}$ play the analogous role for $s = 1$. This defines a bosonic auxiliary Fock space, $\mathcal{H}_{\text{AF}}$, which is spanned by occupation-number states encoding the ADO indices, $|\mathbf{m},\mathbf{n},\mathbf{o},\mathbf{p}\rangle$. Further details of the auxiliary Fock-space structure are given in Appendix~\ref{app: Details of the Auxiliary Fock Space}.

In a similar manner to other Fock-space representations of the hierarchy \cite{Ke2022_hierarchical,Takahashi2024_tensor-train}, this allows one to express the set of ADOs defining the total state as an extended density operator,
\begin{align}
    \left[\rho_{\text{S}},\dots,\rho_{\substack{\mathbf{m}|\mathbf{n}\\\mathbf{o}|\mathbf{p}}},\dots\right] \rightarrow |\rho_{\text{ext}}\rangle\rangle.
\end{align}
Here, $|\rho_{\text{ext}}\rangle\rangle$ is a vector in the auxiliary Fock space whose components are operators acting on the system Hilbert space $\mathcal{H}_{\mathrm{S}}$. Accordingly,
$
|\rho_{\text{ext}}\rangle\rangle \in \mathcal{H}_{O_\mathrm{S}} \otimes \mathcal{H}_{\text{AF}},
$
where $\mathcal{H}_{O_\mathrm{S}}$ denotes the space of linear operators acting on $\mathcal{H}_{\mathrm{S}}$:
\begin{align}
    |\rho_{\text{ext}}\rangle\rangle = \: & \sum_{\substack{\mathbf{m}|\mathbf{n}\\ \mathbf{o}|\mathbf{p}}} \left(\prod_{\ell}\frac{1}{\sqrt{m_{\ell}!\: n_{\ell}!\:o_{\ell}!\:p_{\ell}!}}\right)\rho_{\substack{\mathbf{m}|\mathbf{n}\\ \mathbf{o}|\mathbf{p}}}  \otimes |\mathbf{m},\mathbf{n},\mathbf{o},\mathbf{p}\rangle .
\end{align}
One can then obtain all ADOs via projection of $|\rho_{\text{ext}}\rangle\rangle$,
\begin{align}
    \rho_{\substack{\mathbf{m}|\mathbf{n}\\ \mathbf{o}|\mathbf{p}}} = \: & \langle \langle \mathbf{m},\mathbf{n},\mathbf{o},\mathbf{p}|\rho_{\text{ext}} \rangle\rangle\prod_{\ell}\sqrt{m_{\ell}!\: n_{\ell}!\:o_{\ell}!\:p_{\ell}!},
\end{align}
where $| \mathbf{m},\mathbf{n},\mathbf{o},\mathbf{p} \rangle \rangle = |\mathbf{m},\mathbf{n},\mathbf{o},\mathbf{p} \rangle \otimes \mathbb{I}_{\text{S}}$ and $\mathbb{I}_{\text{S}}$ is the identity operator in $\mathcal{H}_{O_\mathrm{S}}$. For example, the system density matrix is obtained via $\rho_{\text{S}} = \langle\langle \mathbf{0},\mathbf{0},\mathbf{0},\mathbf{0} |\rho_{\text{ext}} \rangle\rangle$.

In this bosonic representation of the ADO index structure, the hierarchy is generated by a linear operator acting on $\mathcal{H}_{O_\mathrm{S}} \otimes \mathcal{H}_{\text{AF}}$, 
\begin{align}
    \frac{\partial}{\partial t} |\rho_{\text{ext}}\rangle\rangle = \: & \mathcal{L} |\rho_{\text{ext}}\rangle\rangle \label{eq: EOM extended space},
\end{align}
where the generator of the extended-space dynamics is expressed in terms of creation and annihilation operators on the auxiliary Fock space and superoperators on the system space, 
  \begin{align}
    \mathcal{L} = \: & -i\mathcal{L}_{\text{S}}\otimes \mathbb{I}_{\text{AF}} - \nonumber \\
    & \sum_{\ell = 1}^{N^{\text{max}}_{\ell}}\gamma_{\ell,r} \mathbb{I}_{\text{S}} \otimes \left(a^{\dag}_{\ell}a^{\pd}_{\ell} + b^{\dag}_{\ell}b^{\pd}_{\ell} + c^{\dag}_{\ell}c^{\pd}_{\ell} + d^{\dag}_{\ell}d^{\pd}_{\ell}\right) - \nonumber \\
    & i \sum_{\ell = 1}^{N^{\text{max}}_{\ell}} \mathcal{L}_{O_{S}} \otimes \left(a^{\pd}_{\ell} + b^{\pd}_{\ell} + c^{\pd}_{\ell} + d^{\pd}_{\ell} \right) - \nonumber \\
    & i \sum_{\ell = 1}^{N^{\text{max}}_{\ell}} \left(\eta^{\pd}_{\ell}O_{S}^{L} \otimes a^{\dag}_{\ell} - \eta^{*}_{\ell} O_{S}^{R} \otimes c^{\dag}_{\ell}\right) - \nonumber \\
    & i \sum_{\ell = 1}^{N^{\text{max}}_{\ell}} \gamma_{\ell,i} \mathbb{I}_{\text{S}} \otimes \left(a^{\dag}_{\ell}b^{\pd}_{\ell} + a^{\pd}_{\ell}b^{\dag}_{\ell} - c^{\dag}_{\ell}d^{\pd}_{\ell} - c^{\pd}_{\ell}d^{\dag}_{\ell}\right). \label{eq: HEOM extended space}
\end{align} 
Here, $\mathcal{L}_{O_{S}} = [O_{S},\cdot]$ defines a commutator of operators in the system space with $O_{S}$ and $O_{S}^{L,R}$ refers to the action of $O_{S}$ on the left and right, respectively. 

The corresponding equations of motion for the individual ADOs are obtained by projecting Eq.\eqref{eq: EOM extended space} onto the auxiliary Fock states $|\mathbf{m},\mathbf{n},\mathbf{o},\mathbf{p} \rangle\rangle$:
  \begin{align}
    \partial_{t}\rho_{\substack{\mathbf{m}|\mathbf{n}\\\mathbf{o}|\mathbf{p}}} = \: & \langle\langle \mathbf{m},\mathbf{n},\mathbf{o},\mathbf{p} | \mathcal{L} | \rho_{\text{ext}} \rangle\rangle \\
= \: & - \left(i\mathcal{L}_{\text{S}} + \sum_{\ell = 1}^{N^{\text{max}}_{\ell}}\gamma_{\ell,r}\left(m_{\ell} + n_{\ell} + o_{\ell} + p_{\ell}\right) \right)\rho_{\substack{\mathbf{m}|\mathbf{n}\\\mathbf{o}|\mathbf{p}}} \nonumber \\ 
    & -i \sum_{\ell} \left[O_{S}, \rho_{\substack{\mathbf{m}^{+}_{\ell}|\mathbf{n} \\ \mathbf{o} |\mathbf{p}}}
	+\rho_{\substack{\mathbf{m}|\mathbf{n}^{+}_{\ell} \\ \mathbf{o} |\mathbf{p}}}
	+\rho_{\substack{\mathbf{m}|\mathbf{n} \\ \mathbf{o}^{+}_{\ell}|\mathbf{p}}}
	+\rho_{\substack{\mathbf{m}|\mathbf{n} \\ \mathbf{o} |\mathbf{p}^{+}_{\ell}}} \right] \nonumber \\
	& -i \sum_{\ell} \left[m_{\ell}\eta_{\ell} O_{S} \rho_{\substack{\mathbf{m}^{-}_{\ell}|\mathbf{n} \\ \mathbf{o} |\mathbf{p}}} 
	-o_{\ell}\eta_{\ell}^*  \rho_{\substack{\mathbf{m}|\mathbf{n} \\ \mathbf{o} ^{-}_{\ell}|\mathbf{p}}}O_{S} \right] \nonumber \\
	& -i \sum_{\ell} \gamma_{\ell,i}
	\left[
	m_{\ell}\rho_{\substack{\mathbf{m}^{-}_{\ell}|\mathbf{n}^{+}_{\ell} \\ \mathbf{o}|\mathbf{p}}} 
	+n_{\ell}\rho_{\substack{\mathbf{m}^{+}_{\ell}|\mathbf{n}^{-}_{\ell} \\ \mathbf{o}|\mathbf{p}}} - \right. \nonumber \\
	& \left. \qquad \qquad \qquad o_{\ell}\rho_{\substack{\mathbf{m}|\mathbf{n} \\ \mathbf{o}^{-}_{\ell}|\mathbf{p}^{+}_{\ell}}}
	-p_{\ell}\rho_{\substack{\mathbf{m}|\mathbf{n}  \\ \mathbf{o}^{+}_{\ell}|\mathbf{p}^{-}_{\ell}}}
	\right]. \label{eq: occ number representation HEOM}
\end{align} 

Next, we will demonstrate that the stability of Eq.\eqref{eq: occ number representation HEOM} under truncation can be directly analyzed from its structure. Since this formulation is isomorphic to Eq.\eqref{eq: standard HEOM}, this offers immediate insight into the origin of numerical instabilities in the HEOM approach in general. Inspecting the generator in Eq.~\eqref{eq: HEOM extended space}, one sees that the system component $\mathcal{L}_{\text{S}}$ acts only within the space of system operators, generating the Hamiltonian part of the reduced dynamics. By contrast, the term proportional to $\gamma_{\ell,i}$ acts within the auxiliary Fock space by mixing the cosine and sine sectors associated with each complex bath-correlation exponent. Taken in isolation, neither of these terms produces monotonic growth of the auxiliary occupation number. The dominant occupation-growth mechanism instead arises from the competition between the damping terms proportional to $\gamma_{\ell,r}$, which suppress large auxiliary occupations, and the hierarchy-raising terms proportional to $\eta_\ell$, which transport weight towards higher tiers. As we show in the following, it is the imbalance between these two mechanisms in the presence of a finite truncation that gives rise to numerical instabilities in the HEOM.


To isolate the mechanism of instability, we now consider a minimal single-direction model obtained by retaining only the damping of one auxiliary occupation and one creation-like hierarchy-raising term. All system-space dynamics, cosine-sine mixing terms associated with $\gamma_{\ell,i}$, and hierarchy-lowering terms are omitted. The resulting generator is
\begin{align}
	L = -\gamma a^\dagger a + \eta a^\dagger, \label{eq: reduced generator}
\end{align}
with $\gamma > 0$. This model is not intended to reproduce the full HEOM dynamics, but to isolate the algebraic structure responsible for large non-normal, transient amplification: a damped auxiliary ladder driven by an unbalanced raising operator.

Here, $a^\dagger$ and $a$ are bosonic creation and annihilation operators acting on an infinite-dimensional Fock space with the standard relations, while $\gamma \in \mathbb{R}^{+}$ and $\eta \in \mathbb{C}$. In practical HEOM calculations, one must introduce a finite truncation tier, which in this formulation corresponds to projecting $L$ onto the finite subspace spanned by $\{|n\rangle\}_{n=0}^{N}$, yielding the truncated operator 
\begin{align}
	L_N	= \: & 	P_N L P_N,	\label{eq: truncated transformed toy generator}
\end{align}
where 
\begin{align}
	P_N=\sum_{n=0}^{N}| n \rangle \langle n |.
\end{align}

For this finite-dimensional restriction, the corresponding spectrum of the truncated generator is purely real and given by
\begin{align}
	\sigma(L_N) = \{-\gamma n \mid n = 0,\dots,N\}. \label{eq: toy spectrum}
\end{align} 
However, to characterize the stability of $L_{N}$, we will consider the time-dependent Euclidean norm of the resulting propagator, 
\begin{align}
    G_{N}(t) = \: & \|e^{L_{N}t}\|_{2}
\end{align} 
where $\| \cdot \|_{2}$ denotes the largest singular value. 

For normal operators satisfying $[A,A^\dagger]=0$, the spectral theorem provides a direct link between spectrum and dynamics:
\begin{align}
	\|e^{tA}\|_{2} = \: & \max_{\lambda \in \sigma(A)} |e^{t\lambda}| = e^{t \max \mathrm{Re}\,\sigma(A)}.
\end{align}
Thus, spectral stability ($\max \mathrm{Re}\,\sigma(A)\le 0$) implies bounded dynamical evolution. However, the HEOM generator is non-normal, and the norm of the propagator is no longer determined solely by the spectrum. In this case, strong finite-time amplification can occur even when all eigenvalues have non-positive real parts.

To understand these transient effects, we employ the Dunford-Cauchy representation of the propagator,
\begin{align}
	e^{tA} = \: & \frac{1}{2\pi i} \oint_\Gamma e^{tz} (z - A)^{-1} dz,
\end{align}
As we show in Appendix~\ref{app: Stability Analysis of Simplified Model}, this form can be used to obtain an estimate of the lower bound of the time-dependent Euclidean norm of the truncated propagator,
\begin{align}
	\sup_{t\geq 0}\|e^{tL_N}\|_{2} \gtrsim \: & \frac{(|\eta|/\gamma)^N}{\sqrt{N!}}. \label{eq: structural insight main text}
\end{align}
Consequently, despite a bounded, stable spectrum, the truncated generator can exhibit very large transient, non-normal amplification. Eq.\eqref{eq: structural insight main text} demonstrates that the onset of these transient instabilities is controlled by the ratio $|\eta|/\gamma$, and is therefore most pronounced in regimes of strong system-bath coupling $(|\eta| \gg  1)$ or slowly decaying bath-correlations $(|\gamma| \ll  1)$, as in the case of highly structured spectral densities and strongly non-Markovian environments.

This simple example therefore illustrates how repeated hierarchy raising along a finite bosonic occupation chain can generate strong amplification even in the presence of a spectrally stable generator. Finite truncation removes the couplings to occupations beyond the highest retained tier and therefore changes the operator structure at the truncation boundary. Note that in the full HEOM in Eq.\eqref{eq: occ number representation HEOM}, hierarchy-raising and hierarchy-lowering couplings coexist along with other rotation terms omitted here, such that the spectrum is not even guaranteed to be bounded under truncation. While these may lead to fundamental long-time instabilities, the approach in this paper is focused on minimizing the transient, non-normal amplification that arises due to truncation of the unpaired hierarchy-raising term $\eta a^{\dag}$.

Consequently, we introduce a Gaussian similarity transformation to the generator 
\begin{align}
    \tilde{L} = e^{-S} L e^{S},
\end{align}
with the motivation that the Gaussian weighting operator, 
\begin{align}
	e^{-S} = \: & \exp\left(-\frac{(a+a^\dagger)^2}{4}\right),
\end{align}
contracts the relevant part of the Fock space in which the dynamics occurs. This can be seen in the explicit transformed form, obtained via the Baker-Campbell-Hausdorff formula, 
\begin{align}
	\tilde{L} = \: & -\frac{\gamma}{4} \left[3a^\dagger a-3a^2 + (a^\dagger)^2 -aa^\dagger \right]
	+	\frac{\eta}{2}	\left(a^\dagger-a\right),	\label{eq: transformed reduced generator}
\end{align}
where the purely hierarchy-raising contribution is transformed into a balanced combination of raising and lowering terms: 
\begin{align}
	\eta a^\dagger \longrightarrow \frac{\eta}{2}\left(a^\dagger-a\right),
\end{align}

To quantify how this change of representation affects stability of the resulting dynamics, we truncate the transformed generator,
\begin{align}
    \tilde{L}_{N} = \: & P_{N} \tilde{L} P_{N}
\end{align}
and compare the Euclidean norm of the resulting propagator
\begin{align}
    \tilde{G}_{N}(t) = \: & \|e^{\tilde{L}_{N}t}\|_{2}
\end{align} 
with the norm of the original propagator. Importantly, $\tilde{L}_{N}$ is not simply related to $L_N$ by a finite-dimensional similarity transformation, since the $P_{N}$ and $e^{-S}$ do not commute.

\begin{figure}
	\begin{center}
		\includegraphics[width=\columnwidth]{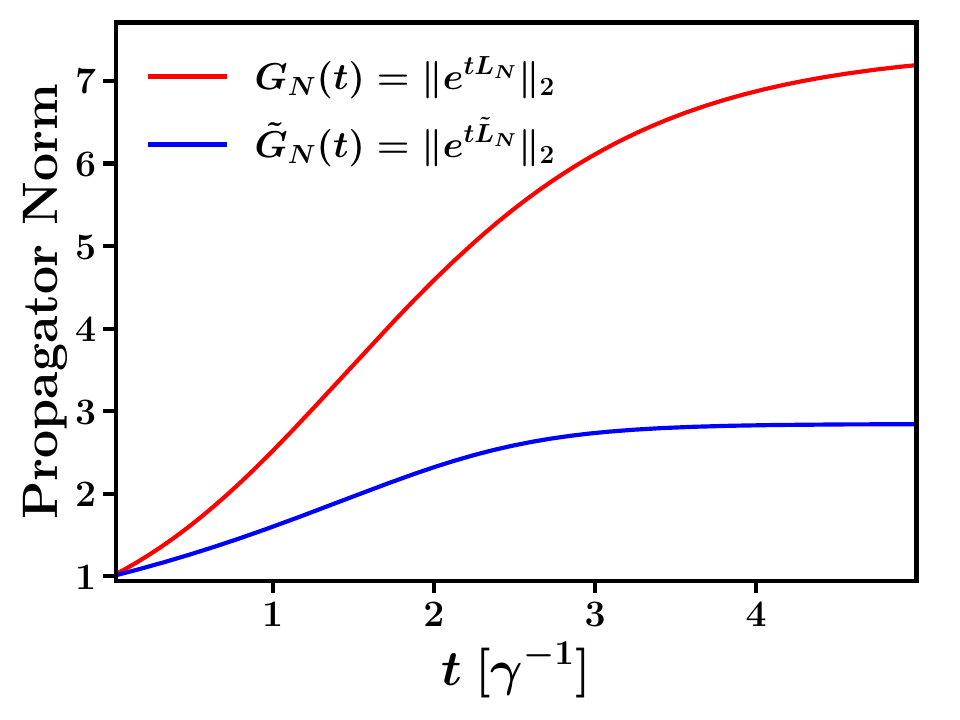}
		\caption{Time-dependent Euclidean norm of the propagator corresponding to the minimal generator, $G_N(t)=\|e^{tL_N}\|_2$, and to the transformed generator, $\tilde{G}_N(t)=\|e^{t\tilde{L}_N}\|_2$. Parameters: $N=50$, $\gamma = 10^{-4}$, and $\eta = 2\times 10^{-4}\cdot(1 + i)$.}
		\label{fig: toy propagator norm}
	\end{center}
\end{figure}

As shown in Fig.~\figref{fig: toy propagator norm}, the standard truncated generator develops a pronounced amplification of the propagator norm despite the absence of eigenvalues with a positive real part. 
In contrast, the Gaussian-transformed generator exhibits a substantially smaller maximum amplification. Parameters are representative of the scale of weights and frequencies obtained in the pole decomposition of the structured spectral density in Fig.~\figref{fig: spectral functions}, with $|\eta|/\gamma = 2.82$. This directly demonstrates that within the minimal model, converting the one-way hierarchy-raising term into a balanced raising-lowering structure suppresses the non-normal amplification associated with finite truncation, and provides the motivation for the stability improvement to the HEOM approach discussed in the next section. 

Finally in this subsection, we note that this amplification mechanism is specific to bosonic HEOM, as fermionic hierarchies terminate at finite tier, preventing the unbounded auxiliary occupation ladder that underlies the instability identified here \cite{Han2018_exact_fermionic}. Although numerical instabilities have been reported when fermionic hierarchies are truncated prematurely using approximate termination schemes \cite{Zhang2021_adiabatic_fermionic}, these are distinct from the finite-truncation instability of the intrinsically infinite bosonic hierarchy considered in this work. Furthermore, although we have analyzed these instabilities in an auxiliary Fock-space picture, they can also be observed in a coordinate representation of the auxiliary space \cite{Ikeda2020_generalization,Li2022_a_low-temperature}, which we discuss in further detail in Appendix~\ref{app: Perspective on Instability in the Coordinate Representation}.

\subsection{Stability Improvement} \label{subsec: Stability Improvement}

Motivated by the analysis of the reduced model in Eq.\eqref{eq: reduced generator}, we now apply the corresponding Gaussian similarity transformation to the full auxiliary Fock space by introducing a Gaussian-weighted extended density operator,
\begin{align}
	|\tilde{\rho}_{\text{ext}}\rangle\rangle
	= \: & \mathcal{N}_{0}	\left(e^{-S} \otimes \mathbb{I}_{\text{S}}\right)	|\rho_{\text{ext}}\rangle\rangle.
\end{align}
Here, $e^{-S}$ acts exclusively on the auxiliary Fock space and is defined in terms of the auxiliary creation and annihilation operators as
\begin{align}
	e^{-S} = \: & \prod_{\ell} \prod_{\alpha \in \left\{ a,b,c,d \right\}} e^{-\frac{\left(\alpha^{\pd}_{\ell} +\alpha^{\dag}_{\ell}\right)^{2}}{4}} \: .	\label{eq: Gaussian weight operator form}
\end{align}

The matrix elements of this operator in the coordinate basis are identical to those appearing in the overlaps between the Fock and coordinate representations; see Eq.\eqref{eq: overlap identity}, for example. The normalization factor $\mathcal{N}_{0}=2^{N^{\text{max}}_{\ell}}$
is chosen such that the Gaussian transformation maps the original
auxiliary vacuum onto the normalized Bogoliubov vacuum. 

In this Gaussian-weighted rescaling, projection onto the original auxiliary Fock states yields the Gaussian-weighted ADOs, which are linear combinations of the original ADOs:
\begin{align}
    \tilde{\rho}_{\substack{\mathbf{m}|\mathbf{n}\\ \mathbf{o}|\mathbf{p}}} = \: & \langle\langle \mathbf{m},\mathbf{n},\mathbf{o},\mathbf{p} | \tilde{\rho}_{\text{ext}}\rangle\rangle \nonumber \\
    = \: & \sum_{\substack{\mathbf{m}',\mathbf{n}',\\ \mathbf{o}',\mathbf{p}'}} \mathsf{\Theta}_{\substack{\mathbf{m}|\mathbf{n}\\ \mathbf{o}|\mathbf{p}};\substack{\mathbf{m}',\mathbf{n}',\\ \mathbf{o}',\mathbf{p}'}} \rho_{\substack{\mathbf{m}',\mathbf{n}',\\ \mathbf{o}',\mathbf{p}'}},
\end{align}
where the coefficients are given by 
\begin{align}
	\mathsf{\Theta}_{
		\substack{\mathbf{m}|\mathbf{n}\\\mathbf{o}|\mathbf{p}};
		\substack{\mathbf{m}'|\mathbf{n}'\\\mathbf{o}'|\mathbf{p}'}
	}
	={}&
	\mathcal{N}_{0}
	\prod_{\ell}
	\sqrt{
		\frac{
			m_{\ell}!n_{\ell}!o_{\ell}!p_{\ell}!
		}{
			m'_{\ell}!n'_{\ell}!o'_{\ell}!p'_{\ell}!
		}
	}
	\nonumber\\
	&\times
	\langle
	\mathbf{m},\mathbf{n},\mathbf{o},\mathbf{p}
	|e^{-S}|
	\mathbf{m}',\mathbf{n}',\mathbf{o}',\mathbf{p}'
	\rangle .
\end{align}
Consequently, in this picture, the Gaussian-weighted extended state can be written as
\begin{align}
    |\tilde{\rho}_{\text{ext}} \rangle\rangle = \: & \sum_{\substack{\mathbf{m}|\mathbf{n}\\ \mathbf{o}|\mathbf{p}}} \left(\prod_{\ell}\frac{1}{\sqrt{m_{\ell}!\: n_{\ell}!\:o_{\ell}!\:p_{\ell}!}}\right) \tilde{\rho}_{\substack{\mathbf{m}|\mathbf{n}\\ \mathbf{o}|\mathbf{p}}} \otimes | \mathbf{m},\mathbf{n},\mathbf{o},\mathbf{p} \rangle,
\end{align}
which represents a reweighted rather than a new hierarchy. 
The corresponding equation of motion for $|\tilde{\rho}_{\text{ext}}\rangle\rangle$ is 
  \begin{align}
    \partial_{t} |\tilde{\rho}_{\text{ext}} \rangle\rangle = \: & \mathcal{N}_{0}e^{-S} \mathcal{L} | \rho_{\text{ext}} \rangle\rangle \\
    = \: & \mathcal{N}_{0} e^{-S} \mathcal{L}e^{S}e^{-S} | \rho_{\text{ext}}\rangle\rangle \\
    = \: & \tilde{\mathcal{L}} |\tilde{\rho}_{\text{ext}}\rangle\rangle \label{eq: similarity transformed HEOM},
\end{align} 
with $\tilde{\mathcal{L}} = e^{-S} \mathcal{L}e^{S}$. Next, we evaluate the rescaled generator $\tilde{\mathcal{L}}$ using Eq.\eqref{eq: Gaussian weight operator form} and the Baker-Campbell-Hausdorff formula,
  \begin{align}
    \tilde{\mathcal{L}} = & -i\mathcal{L}_{\text{S}}  \nonumber \\
    & -\sum_{\ell = 1}^{N^{\text{max}}_{\ell}}\frac{\gamma_{\ell,r}}{4} \left[ \sum_{\alpha \in\{a,b,c,d\}} \hspace{-4mm}\left(3\alpha^{\dag}_{\ell}\alpha^{\pd}_{\ell} - 3\alpha^{\pd}_{\ell}\alpha^{\pd}_{\ell} + \alpha^{\dag}_{\ell}\alpha^{\dag}_{\ell}  - \alpha^{\pd}_{\ell}\alpha^{\dag}_{\ell}\right)\right]  \nonumber \\
    & -\frac{i}{2} \sum_{\ell = 1}^{N^{\text{max}}_{\ell}} \mathcal{L}_{O_{S}} \sum_{\alpha \in\{a,b,c,d\}} \left(3 \alpha^{\pd}_{\ell} + \alpha^{\dag}_{\ell}\right) \nonumber \\
    & -\frac{i}{2} \sum_{\ell = 1}^{N^{\text{max}}_{\ell}} \left(\eta^{\pd}_{\ell}O_{S}^{L} \left(a^{\dag}_{\ell} - a^{\pd}_{\ell}\right) - \eta^{*}_{\ell} O_{S}^{R} \left(c^{\dag}_{\ell} - c^{\pd}_{\ell}\right)\right)  \nonumber \\
    & -i \sum_{\ell = 1}^{N^{\text{max}}_{\ell}} \frac{\gamma_{\ell,i}}{2} \left(a^{\dag}_{\ell}b^{\pd}_{\ell} - 3a^{\pd}_{\ell}b^{\pd}_{\ell} + a^{\dag}_{\ell}b^{\dag}_{\ell} + a^{\pd}_{\ell}  b^{\dag}_{\ell}\right)   \nonumber \\
    & +i \sum_{\ell = 1}^{N^{\text{max}}_{\ell}} \frac{\gamma_{\ell,i}}{2} \left(c^{\dag}_{\ell}d^{\pd}_{\ell} - 3c^{\pd}_{\ell}d^{\pd}_{\ell} + c^{\dag}_{\ell}d^{\dag}_{\ell} + c^{\pd}_{\ell} d^{\dag}_{\ell}\right). \label{eq: HEOM extended space transformed 1}
\end{align} 
Note that we have suppressed the identity operators and tensor notation for brevity. While the similarity transformation leaves the spectrum of the untruncated generator invariant, it changes the finite auxiliary representation on which truncation is imposed. As a result, the truncated transformed generator can exhibit substantially reduced pseudospectral spreading and weaker transient amplification than the corresponding standard HEOM truncation.

To make the stabilizing structure introduced by the Gaussian similarity transformation explicit at the operator level, we perform a further reorganization of the auxiliary Fock-space generators in terms of rotated ladder operators that align the dominant dissipative directions with the occupation basis.
Using a Bogoliubov-like transformation, we define the new operators 
\begin{align}
	A^{\pd}_{\ell} & = \frac{1}{2\sqrt{2}}\left( 3a^{\pd}_{\ell}+a^\dagger_{\ell} \right) , \qquad A_{\ell}^\dagger= \frac{1}{2\sqrt{2}}\left( 3a^\dagger_{\ell}+a^{\pd}_{\ell} \right)\nonumber\\
	B^{\pd}_{\ell} & = \frac{1}{2\sqrt{2}}\left( 3b^{\pd}_{\ell}+b^\dagger_{\ell} \right) , \qquad B_{\ell}^\dagger= \frac{1}{2\sqrt{2}}\left( 3b^\dagger_{\ell}+b^{\pd}_{\ell} \right)\nonumber\\
	C^{\pd}_{\ell} & = \frac{1}{2\sqrt{2}}\left( 3c^{\pd}_{\ell}+c^\dagger_{\ell} \right) , \qquad C_{\ell}^\dagger= \frac{1}{2\sqrt{2}}\left( 3c^\dagger_{\ell}+c^{\pd}_{\ell} \right)\nonumber\\
	D^{\pd}_{\ell} & = \frac{1}{2\sqrt{2}}\left( 3d^{\pd}_{\ell}+d^\dagger_{\ell} \right) , \qquad D_{\ell}^\dagger= \frac{1}{2\sqrt{2}}\left( 3d^\dagger_{\ell}+d^{\pd}_{\ell} \right), \label{eq: Bog operators}
\end{align}
which can be used to rewrite the HEOM generator $\tilde{\mathcal{L}}$, as shown in Eq.\eqref{eq: HEOM extended space transformed 2} in Appendix~\ref{app: Coordinate Representation of Bogoliubov-Transformed Operators}.

Now, the operators $A_{\ell}^{(\dag)}$, $B_{\ell}^{(\dag)}$, $C_{\ell}^{(\dag)}$, and $D_{\ell}^{(\dag)}$ define valid bosonic modes, as they preserve the canonical commutation relations,
\begin{align}
    \left[A^{\pd}_{\ell},A^{\dag}_{\ell}\right] = \left[B^{\pd}_{\ell},B^{\dag}_{\ell}\right] = \left[C^{\pd}_{\ell},C^{\dag}_{\ell}\right] = \left[D^{\pd}_{\ell},D^{\dag}_{\ell}\right] = 1.
\end{align}
As shown in Appendix~\ref{app: Coordinate Representation of Bogoliubov-Transformed Operators}, these operators generate excitations in the transformed auxiliary Fock space spanned by the occupation-number basis $|\mathbf{M},\mathbf{N},\mathbf{O},\mathbf{P}\rangle$. In this representation, the Gaussian-weighted extended density operator can be expanded as 
\begin{align}
    |\tilde{\rho}_{\text{ext}} \rangle\rangle = \: & \sum_{\substack{\mathbf{M}|\mathbf{N}\\ \mathbf{O}|\mathbf{P}}} \left(\prod_{\ell} \frac{1}{\sqrt{M_{\ell}!\: N_{\ell}!\:O_{\ell}!\:P_{\ell}!}}\right)\tilde{\rho}_{\substack{\mathbf{M}|\mathbf{N}\\ \mathbf{O}|\mathbf{P}}}  \otimes |\mathbf{M},\mathbf{N},\mathbf{O},\mathbf{P}\rangle,
\end{align}
and the equations of motion for the $\tilde{\rho}_{\substack{\mathbf{M}|\mathbf{N}\\ \mathbf{O}|\mathbf{P}}}$ are obtained by projecting Eq.\eqref{eq: similarity transformed HEOM}, in the form of Eq.\eqref{eq: HEOM extended space transformed 2}, onto $|\mathbf{M},\mathbf{N},\mathbf{O},\mathbf{P}\rangle\rangle$:
  \begin{align}
	& \partial_{t}\tilde{\rho}_{\substack{\mathbf{M}|\mathbf{N}\\\mathbf{O}|\mathbf{P}}} = \: \langle\langle \mathbf{M},\mathbf{N},\mathbf{O},\mathbf{P} | \tilde{\mathcal{L}} | \tilde{\rho}_{\text{ext}} \rangle\rangle \\
	& \qquad = \: - \left(i\mathcal{L}_{\text{S}} + \sum_{\ell = 1}^{N^{\text{max}}_{\ell}}\gamma_{\ell,r}\left(M_{\ell} + N_{\ell} + O_{\ell} + P_{\ell}\right) \right)\tilde{\rho}_{\substack{\mathbf{M}|\mathbf{N}\\\mathbf{O}|\mathbf{P}}} \nonumber \\ 
	& \qquad + \sum_{\ell = 1}^{N^{\text{max}}_{\ell}}\gamma_{\ell,r}\left(\tilde{\rho}_{\substack{\mathbf{M}_{\ell}^{+2}|\mathbf{N}\\\mathbf{O}|\mathbf{P}}} + \tilde{\rho}_{\substack{\mathbf{M}|\mathbf{N}_{\ell}^{+2}\\\mathbf{O}|\mathbf{P}}} + \tilde{\rho}_{\substack{\mathbf{M}|\mathbf{N}\\\mathbf{O}_{\ell}^{+2}|\mathbf{P}}} + \tilde{\rho}_{\substack{\mathbf{M}|\mathbf{N}\\\mathbf{O}|\mathbf{P}_{\ell}^{+2}}}\right) \nonumber \\
	& \qquad - \sqrt{2} i \sum_{\ell} \mathcal{L}_{O_{S}} \left(\tilde{\rho}_{\substack{\mathbf{M}^{+}_{\ell}|\mathbf{N} \\ \mathbf{O} |\mathbf{P}}}
	+\tilde{\rho}_{\substack{\mathbf{M}|\mathbf{N}^{+}_{\ell} \\ \mathbf{O} |\mathbf{P}}}
	+\tilde{\rho}_{\substack{\mathbf{M}|\mathbf{N} \\ \mathbf{O}^{+}_{\ell}|\mathbf{P}}}
	+\tilde{\rho}_{\substack{\mathbf{M}|\mathbf{N} \\ \mathbf{O} |\mathbf{P}^{+}_{\ell}}} \right) \nonumber \\
	& \qquad -\frac{i}{\sqrt{2}} \sum_{\ell} \left[\eta_{\ell} O_{S}^{L} \left(M_{\ell}\tilde{\rho}_{\substack{\mathbf{M}^{-}_{\ell}|\mathbf{N} \\ \mathbf{O} |\mathbf{P}}} - \tilde{\rho}_{\substack{\mathbf{M}^{+}_{\ell}|\mathbf{N} \\ \mathbf{O} |\mathbf{P}}}\right) - \right. \nonumber \\
	& \qquad \qquad \qquad \:\:\:\: \left. \eta_{\ell}^*  \left(O_{\ell}\tilde{\rho}_{\substack{\mathbf{M}|\mathbf{N} \\ \mathbf{O}^{-}_{\ell}|\mathbf{P}}} - \tilde{\rho}_{\substack{\mathbf{M}|\mathbf{N} \\ \mathbf{O}^{+}_{\ell}|\mathbf{P}}}\right)O_{S}^{R} \right] \nonumber \\
	& \qquad -i \sum_{\ell} \gamma_{\ell,i}
	\left[M_{\ell}\tilde{\rho}_{\substack{\mathbf{M}^{-}_{\ell}|\mathbf{N}^{+}_{\ell} \\ \mathbf{O}|\mathbf{P}}} 
	-2\tilde{\rho}_{\substack{\mathbf{M}^{+}_{\ell}|\mathbf{N}^{+}_{\ell} \\ \mathbf{O}|\mathbf{P}}} + N_{\ell}\tilde{\rho}_{\substack{\mathbf{M}^{+}_{\ell}|\mathbf{N}^{-}_{\ell} \\ \mathbf{O}|\mathbf{P}}} + \right] \nonumber \\
	& \qquad + i \sum_{\ell} \gamma_{\ell,i}
	\left[O_{\ell}\tilde{\rho}_{\substack{\mathbf{M}|\mathbf{N} \\ \mathbf{O}^{-}_{\ell}|\mathbf{P}^{+}_{\ell}}} 
	-2\tilde{\rho}_{\substack{\mathbf{M}|\mathbf{N} \\ \mathbf{O}^{+}_{\ell}|\mathbf{P}^{+}_{\ell}}} + P_{\ell}\tilde{\rho}_{\substack{\mathbf{M}|\mathbf{N} \\ \mathbf{O}^{+}_{\ell}|\mathbf{P}^{-}_{\ell}}} + \right]. \label{eq: occ number representation final form}
\end{align} 
where $\mathbf{M}^{+n}_{\ell} = (M_{0},\dots,M_{\ell}+n,\dots,M_{N^{\text{max}}_{\ell}})$ and similarly for $\mathbf{N}^{+n}_{\ell}$, $\mathbf{O}^{+n}_{\ell}$, and $\mathbf{P}^{+n}_{\ell}$.

\subsection{Stabilized Algorithm} \label{subsec: Stabilized Algorithm}

The equations of motion given in Eq.~\eqref{eq: occ number representation final form} define a transformed hierarchy in which the finite-truncation instabilities of the original HEOM in Eq.~\eqref{eq: occ number representation HEOM} are substantially suppressed. Consequently, the transformed hierarchy provides a more stable route to computing physical observables.

In practice, the dynamics can be simulated by implementing the following procedure:
\begin{enumerate}
    \item The HEOM approach presented in this work assumes that the initial state factorizes, as shown in Eq.\eqref{eq: initial condition}. In the auxiliary Fock space representation, this corresponds to the vacuum in all hierarchy modes,
    \begin{align}
        |\rho_{\text{ext}}(0) \rangle\rangle = \rho_{\text{S}}(0) \otimes | 0 \rangle, \label{eq: IC fock basis}
    \end{align}
    which then defines the corresponding initial condition in the Gaussian-weighted and Bogoliubov-rotated representation:
    \begin{align}
        \tilde{\rho}_{\substack{\mathbf{M}|\mathbf{N}\\\mathbf{O}|\mathbf{P}}}(0) = \mathcal{N}_{0} \: \langle \mathbf{M},\mathbf{N},\mathbf{O},\mathbf{P} | e^{-S} | 0 \rangle \rho_{\text{S}}(0). \label{eq: IC fock ADOs}
    \end{align}
    It can be shown that for the uncorrelated initial condition given in Eq.\eqref{eq: IC fock basis}, Eq.\eqref{eq: IC fock ADOs} yields zero for all ADOs of nonzero tier, $\tilde{\rho}_{\substack{\mathbf{M}|\mathbf{N}\\\mathbf{O}|\mathbf{P}}}(0) = 0$.

    \item The quantities $\tilde{\rho}_{\substack{\mathbf{M}|\mathbf{N}\\\mathbf{O}|\mathbf{P}}}(t)$ are then evolved according to the stable hierarchy in Eq.\eqref{eq: occ number representation final form}, using a suitable numerical integrator. In this work, an adaptive timestep fourth-order Runge-Kutta approach was used \cite{Ke2022_nonequilibrium}.
    \item Physical observables are recovered by transforming back to the original representation. In particular, the reduced system density operator is obtained as
    \begin{align}
        \rho_{\text{S}}(t) = \: & \mathcal{N}_{0}^{-1} \langle 0 | e^{+S} |\tilde{\rho}_{\text{ext}}(t) \rangle\rangle \\
        = \: & \mathcal{N}_{0}^{-1} \sum_{\substack{\mathbf{M}|\mathbf{N}\\ \mathbf{O}|\mathbf{P}}} \left(\prod_{\ell} \frac{1}{\sqrt{M_{\ell}!\: N_{\ell}!\:O_{\ell}!\:P_{\ell}!}}\right) \times \nonumber \\
        & \qquad\qquad \tilde{\rho}_{\substack{\mathbf{M}|\mathbf{N}\\ \mathbf{O}|\mathbf{P}}}(t) \langle 0 | e^{+S} |\mathbf{M},\mathbf{N},\mathbf{O},\mathbf{P}\rangle. \label{eq: inverse similarity transformation rho_S}
    \end{align}
\end{enumerate}

The required matrix elements $\langle 0| e^{+S} |\mathbf{M},\mathbf{N},\mathbf{O},\mathbf{P}\rangle$ can be analytically evaluated, as shown in Appendix~\ref{app: Similarity Transformation} and Refs.~\cite{Li2022_a_low-temperature,Xu2026_markovian}, which yields the explicit reconstruction formula
\begin{align}
    \rho_{\text{S}}(t) = \: & \sum_{\substack{\mathbf{M}|\mathbf{N}\\ \mathbf{O}|\mathbf{P}}}\tilde{\Theta}^{-1}_{\substack{\mathbf{2M}|\mathbf{2N}\\ \mathbf{2O}|\mathbf{2P}}}\tilde{\rho}_{\substack{\mathbf{2M}|\mathbf{2N}\\ \mathbf{2O}|\mathbf{2P}}}(t),
\end{align}
where 
\begin{align}
     \tilde{\Theta}^{-1}_{\substack{\mathbf{2M}|\mathbf{2N}\\ \mathbf{2O}|\mathbf{2P}}} = \: & \prod_{\ell} \frac{(2M_{\ell} - 1)!!(2N_{\ell} - 1)!!(2O_{\ell} - 1)!!(2P_{\ell} - 1)!!}{(2M_{\ell})!\: (2N_{\ell})!\:(2O_{\ell})!\:(2P_{\ell})!}.
\end{align}

Finally, we note that in this work, we adopt the so-called $L$-truncation scheme \cite{Krug2023_on_stability}. When applying this scheme to the transformed, stabilized HEOM introduced in the previous section, this amounts to excluding all ADOs $\tilde{\rho}_{\substack{\mathbf{M}|\mathbf{N}\\ \mathbf{O}|\mathbf{P}}}$ that satisfy 
\begin{align}
    \sum_{\ell = 1}^{N^{\text{max}}_{\ell}} \left(M_{\ell} + N_{\ell} + O_{\ell} + P_{\ell}\right) > \: & n_{\text{max}},
\end{align}
where $n_{\text{max}}$ is the predetermined hierarchy truncation tier. 

In the following section, we will also provide simulation results for the standard HEOM, given in Eq.\eqref{eq: standard HEOM}, to which we will also apply the $L$-truncation scheme. However, since the standard HEOM does not include an extra ADO index from the cosine and sine splitting, the same truncation tier $n_{\text{max}}$ does not correspond to a term-by-term identical truncation of the influence functional in the two representations. In fact, the same $n_{\text{max}}$ corresponds to an effective higher truncation tier for the standard HEOM, as each ADO index essentially contains a sum over the sine and cosine terms. As we show in the next section, this demonstrates the power of the stabilized approach, as even with an effectively lower tier, it generates generally stabler dynamics.

\section{Results} \label{sec: Results}

\noindent In this section, we assess the stability of the standard and transformed HEOM formulations under truncation at finite tier and for the two spectral densities shown in Fig.~\figref{fig: spectral functions}. The Brownian oscillator spectral density is discussed first in Sec.~\ref{subsec: Brownian Oscillator}, where it provides a controlled benchmark for the onset of truncation-induced instabilities as the system-bath coupling is increased. We then turn in Sec.~\ref{subsec: Structured Spectral Density} to the more demanding case of the structured spectral density. In both instances, we compare the resulting population dynamics between the two approaches and examine how the stability behavior changes with the hierarchy truncation tier, $n_{\mathrm{max}}$.

\subsection{Brownian Oscillator Spectral Density} \label{subsec: Brownian Oscillator}

\begin{figure}
    \begin{center}
       \includegraphics[width=\columnwidth]{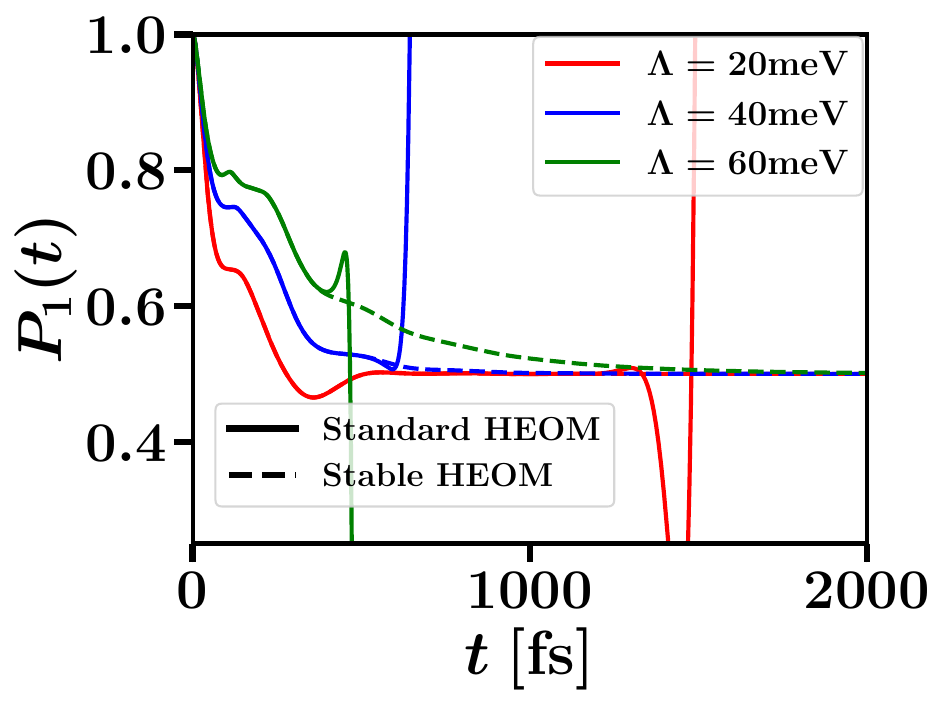}
       \caption{Population of the initially occupied state, $P_{1}(t)=\rho_{\mathrm{S},11}(t)$, as a function of time for the Brownian oscillator spectral density and various reorganization energies, $\Lambda$. Parameters are: $\gamma = 4.9\,\mathrm{meV}$, $\Omega = 13.89\,\mathrm{meV}$, $k_{\mathrm B}T = 25.85\,\mathrm{meV}$, $\varepsilon = 0$, and $\Delta = 10\,\mathrm{meV}$. The truncation tier is chosen such that the stabilized HEOM is converged, which changes for varying reorganization strength: the largest value is $n_{\text{max}} = 14$ for $\Lambda = 60\text{ meV}$.}
       \label{fig: brownian oscillator comparison}
    \end{center}
\end{figure}

In Fig.~\figref{fig: brownian oscillator comparison}, we show the population of the initially occupied state as a function of time, $P_{1}(t)=\rho_{\mathrm{S},11}(t)$, for different system-bath coupling strengths. For this and all subsequent simulations, the initial condition is $\rho_{\mathrm{S},11}(0)=1$, and the spin-boson model is unbiased, $\varepsilon = 0$. Consequently, the two states are equally populated in the stationary regime, 
\begin{align}
    P_{0}(t\rightarrow\infty)=P_{1}(t\rightarrow\infty) = 0.5.
\end{align}
The early-time coherent oscillations are generated by the tunneling term $\Delta\sigma_x$ in $H_{\text{S}}$, while their damping and long-time relaxation are controlled by the coupling to the Brownian oscillator environment.

For the parameters used here, the Brownian oscillator is underdamped, $\gamma\ll\Omega$, and its characteristic frequency is comparable to the intrinsic system timescale set by $\Delta$. The bath therefore induces visible memory effects in the population dynamics, reflected in damped oscillations before relaxation towards the unbiased stationary value. For all values of $\Lambda$ shown, the transformed HEOM follows this relaxation smoothly and remains within the physical population range. By contrast, the standard HEOM eventually becomes unstable: the population leaves the physical interval and diverges after a certain time. Note that as the reorganization energy increases, this occurs earlier in the simulation time. As a result, increasing $\Lambda$ not only increases the damping of the coherent oscillations but also enhances the instability of the standard HEOM approach under finite truncation tier.

\begin{figure}
    \begin{center}
       \includegraphics[width=\columnwidth]{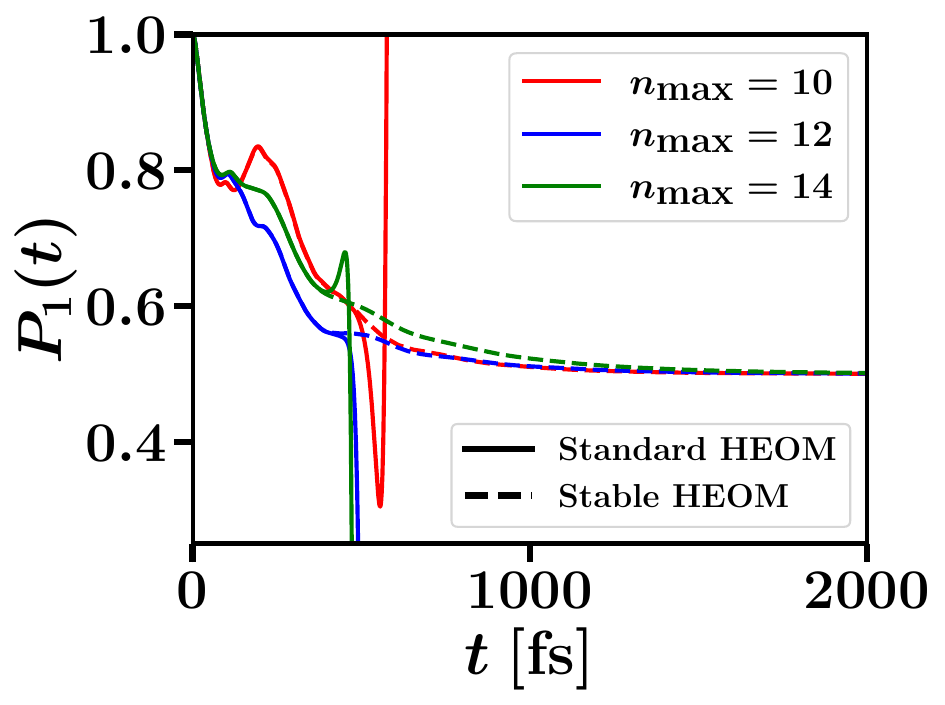}
       \caption{Population of the initially occupied state, $P_{1}(t)=\rho_{\mathrm{S},11}(t)$, as a function of time for the Brownian oscillator spectral density and various truncation tiers, $n_{\text{max}}$. The reorganization energy is $\Lambda = 60\,\mathrm{meV}$, and all other parameters are the same as in Fig.~\figref{fig: brownian oscillator comparison}.}
       \label{fig: brownian oscillator increasing Nmax evidence}
    \end{center}
\end{figure}

In Fig.~\figref{fig: brownian oscillator increasing Nmax evidence}, $P_{1}(t)$ is shown for fixed reorganization energy, $\Lambda=60\,\mathrm{meV}$, while varying the truncation tier $n_{\text{max}}$. The standard HEOM does not show systematic stabilization upon increasing $n_{\text{max}}$. Indeed, as the truncation tier is increased from   $n_{\text{max}} = 10$ to $n_{\text{max}} = 14$, the population dynamics grow unstable at \textit{earlier} simulation times. This counterintuitive behavior is consistent with the non-normal amplification mechanism identified in the minimal model of Sec.~\ref{subsec: Stability Analysis}. Increasing $n_{\text{max}}$ extends the auxiliary occupation chain over which repeated hierarchy-raising processes can act and can therefore enhance the pseudospectral sensitivity of the finite generator. Thus, increasing the hierarchy depth does not necessarily lead to monotonic stabilization of the standard HEOM. In contrast, the transformed HEOM displays a clear convergence pattern, the higher-tier calculations remaining stable and approaching the correct long-time population. This behavior supports the interpretation  in Sec.~\ref{subsec: Stability Analysis}, showing that the instability is not simply an unconverged $n_{\text{max}}$, but originates from transient amplification in the finite, non-normal HEOM generator. Specifically, in the full auxiliary space, the transformed and untransformed generators are related by an invertible transformation. After truncation, however, the two finite generators may not have the same pseudospectral properties, and the transformed representation can therefore display substantially weaker transient amplification.

\subsection{Structured Spectral Density} \label{subsec: Structured Spectral Density}

\begin{figure}
    \begin{center}
       \includegraphics[width=\columnwidth]{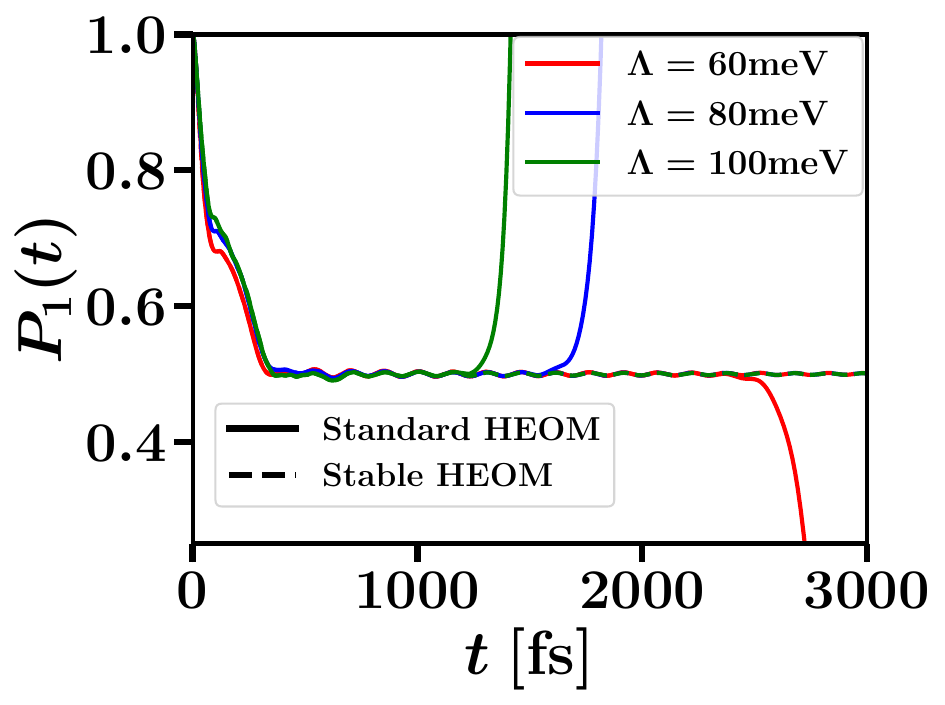}
       \caption{Population dynamics of the initially occupied state, $P_{1}(t)=\rho_{\mathrm{S},11}(t)$, for the structured spectral density shown in Fig.~\figref{fig: spectral functions} and for various reorganization energies, $\Lambda$. Truncation tier is $n_{\text{max}} = 8$ for all reorganization energies. All other parameters are the same as in Fig.~\figref{fig: brownian oscillator comparison}.}
       \label{fig: structured spectral density comparison}
    \end{center}
\end{figure}

Next, in Fig.~\figref{fig: structured spectral density comparison}, we show the population dynamics obtained when the bath is characterized by the structured spectral density shown in Fig.~\figref{fig: spectral functions}. This spectral density contains a broad low-frequency component with characteristic frequencies and linewidths comparable to those of the Brownian oscillator considered above. In addition, it contains narrower spectral features at higher frequencies, in particular around $\omega\simeq 25\text{-}30\,\mathrm{meV}$. Although these features lie above the dominant system timescale set by $\Delta$, they are still thermally accessible at room temperature and contribute to the bath-correlation function through temperature-induced fluctuations.

Compared with the Brownian oscillator case, the structured spectral density is associated with faster relaxation of the initially occupied state towards the unbiased stationary value. The transformed HEOM results show that, after a short transient, the population remains close to $P_{1}=0.5$ over the full propagation window for all couplings considered. The standard HEOM initially follows the same physical relaxation, but eventually becomes unstable. The onset of this instability again moves to earlier times as $\Lambda$ is increased: for $\Lambda=60\,\mathrm{meV}$ the divergence occurs only at late times, whereas for $\Lambda=80\,\mathrm{meV}$ and $100\,\mathrm{meV}$ it appears substantially earlier.

Due to these two distinct components of $J_{\text{Str.}}(\omega)$, the system experiences two competing bath timescales. The broad component provides a continuum of bath modes that efficiently dissipate energy and dephase system coherences on relatively short timescales, closely resembling the dynamics observed for a single Brownian oscillator. In contrast, the narrow  high-frequency feature introduces a long-lived bath mode with a substantially longer memory time, which can exchange coherence with the system over extended periods. As a result, the system no longer exhibits oscillations characterized by a single renormalized frequency; instead, multiple dressed system-bath modes contribute, producing oscillations of reduced amplitude and time-dependent frequency. This behavior is a clear manifestation of non-Markovian dynamics induced by structured environmental coupling and cannot be captured by models based on a single effective bath mode or short-memory approximations.

This additional spectral structure and resulting non-Markovian behavior also has a direct impact on the numerical stability of the finite HEOM truncation. In the standard HEOM, instabilities appear already at moderate coupling strengths,  as signaled by an unphysical growth of $P_{1}(t)$. Stable dynamics can only be simulated with the standard HEOM approach in the very weak-coupling regime $\Lambda \ll k_{\mathrm B}T$. By contrast, the transformed HEOM remains stable over the full propagation window for all couplings shown in Fig.~\figref{fig: structured spectral density comparison}. This indicates that the structured spectral density provides a more stringent stability test than the Brownian oscillator: the larger number of bath-correlation components and the longer-lived memory contributions make the finite hierarchy more susceptible to non-normal transient amplification.

\begin{figure} 
	\begin{center} 
    \includegraphics[width=\columnwidth]{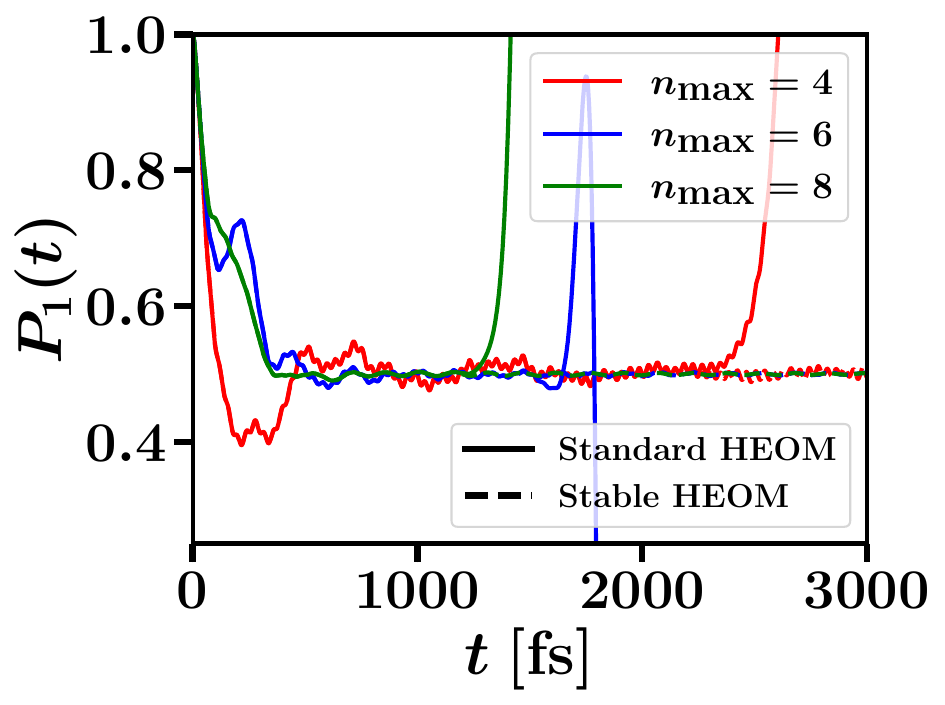} 
    \caption{Population of the initially occupied state, $P_{1}(t)=\rho_{\mathrm{S},11}(t)$, as a function of time for the structured spectral density and various truncation tiers, $n_{\text{max}}$. The reorganization energy is $\Lambda = 100\,\mathrm{meV}$, and all other parameters are the same as in Fig.~\figref{fig: brownian oscillator comparison}. Note that there is extremely close agreement between the stabilized results for all $n_{\text{max}}$, as convergence is achieved quickly for these parameters.} 
    \label{fig: structured spectral density increasing Nmax evidence}
	\end{center} 
\end{figure} 

The tier dependence shown in Fig.~\figref{fig: structured spectral density increasing Nmax evidence} further illustrates this point. For the standard HEOM, increasing $n_{\text{max}}$ does not lead to a monotonic convergence pattern. Instead, the standard HEOM eventually becomes unstable at all tiers shown, with the instability time depending sensitively on the chosen truncation tier. This sensitivity to $n_{\text{max}}$ is characteristic of a finite non-normal generator whose transient amplification is strongly affected by the truncation boundary. The transformed HEOM behaves differently. The curves obtained at different $n_{\text{max}}$ remain stable and collapse towards the same long-time population, indicating convergence of the transformed finite hierarchy. Given that using the same $n_{\text{max}}$ for both approaches induces an effectively larger truncation tier for the standard HEOM, we conclude that the improved stability is  simply a manifestation of the finite-truncation effect in different representations. While the exact untruncated dynamics is unchanged, the transformed finite generator is less susceptible to the transient amplification that destabilizes the standard HEOM.


\section{Conclusion} \label{sec: Conclusion}

In this work, we have analyzed the origin of numerical instabilities within the bosonic hierarchical equations of motion approach and implemented a stabilized formulation that remains robust in strongly coupled and non-Markovian regimes. Via an auxiliary Fock-space representation of the hierarchy, we identified the HEOM generator as a highly non-normal operator whose finite-tier truncation can induce artificial amplification towards increasing auxiliary occupation, resulting in numerical divergences of the dynamics. Utilizing recent insights connecting the HEOM approach to quantum Fokker-Planck formulations and more general extended-space embeddings, we applied a non-unitary similarity transformation acting directly on the auxiliary Fock space, followed by a Bogoliubov-type rotation that reweights and rebalances upward and downward couplings within the hierarchy. In the full auxiliary space, this transformation preserves the underlying exact dynamics. After finite truncation, however, it changes the numerical generator and can substantially suppress transient amplification.

We demonstrated the effectiveness of this approach via two example scenarios. First, we simulated the population dynamics of a spin-boson model coupled to an underdamped Brownian oscillator spectral density. We showed that, although the standard HEOM approach could yield stable dynamics for moderate system-bath couplings, numerical instabilities always appear at some critical reorganization energy. Moreover, we demonstrated that these instabilities cannot be resolved by simply increasing the truncation tier of the hierarchy. In contrast, the transformed HEOM approach yields stable, converged dynamics even in the moderate to strong coupling regime. These results were then further reinforced by investigating the dynamics of a spin-boson model coupled to a structured spectral density. We showed that the additional spectral structure induces significant non-Markovian effects in the population dynamics, causing instabilities in the standard HEOM approach already at weak to moderate system-bath coupling. Again, however, the transformed HEOM approach proved stable for all regimes considered.

Taken together, these findings substantially broaden the range of systems that can be treated reliably using the HEOM approach, particularly in regimes characterized by strong system-bath coupling, structured environments, and pronounced non-Markovian effects. Our results also support the broader view that non-Markovian dynamics can admit multiple equivalent extended-space embeddings whose finite-dimensional truncations possess different numerical stability properties. Within this landscape, the auxiliary Fock-space transformation introduced here provides a stability-oriented representation of bosonic HEOM, particularly useful for structured spectral densities where standard truncations suffer from strong transient amplification. The stabilized hierarchy introduced here enables numerically exact simulations in settings directly relevant to current experiments in quantum technologies, condensed-phase chemistry, and nanoscale materials.

\section*{Acknowledgements}

This work was supported by the Deutsche Forschungsgemeinschaft (DFG) within the framework of the Research Unit FOR5099 “Reducing complexity of nonequilibrium systems”. The authors acknowledge the support by the state of Baden-Württemberg through bwHPC and the DFG through Grant No. INST 40/575-1 FUGG (JUSTUS 2 cluster). E.R. would like to thank the Alexander von Humboldt Foundation for the Humboldt Research Award and the Israel Science Foundation for financial support (Grant No. 4085/25). B.H. acknowledges that much of the work was completed while at the Department of Chemistry in the University of California, Berkeley. Grateful thanks are extended to Riley J. Preston for helpful discussions on the stabilized hierarchy. 

\appendix

\section{Details of the Auxiliary Fock Space} \label{app: Details of the Auxiliary Fock Space}

To formalize the auxiliary occupation-number representation of HEOM in Eq.\eqref{eq: occ number representation HEOM}, we introduce a bosonic representation of the $\mathcal{B}_{(g,\ell,\sigma)}$ functionals by associating to each index $(g,\ell,\sigma)$ a creation operator:
\begin{align}
    \mathcal{B}_{(\cos,\ell,\sigma=0)} = \: & \mathcal{B}_{(\ell,\sigma=0)|} \longrightarrow a_{\ell}^{\dag} \\
    \mathcal{B}_{(\sin,\ell,\sigma=0)} = \: & \mathcal{B}_{|(\ell,\sigma=0)} \longrightarrow b_{\ell}^{\dag} \\
    \mathcal{B}_{(\cos,\ell,\sigma=1)} = \: & \mathcal{B}_{(\ell,\sigma=1)|} \longrightarrow c_{\ell}^{\dag} \\
    \mathcal{B}_{(\sin,\ell,\sigma=1)} = \: & \mathcal{B}_{|(\ell,\sigma=1)} \longrightarrow d_{\ell}^{\dag}, \label{eq: creation operator definition}
\end{align}
which also implies the existence of the corresponding annihilation operators, $a_{\ell}^{\pd}$, $b_{\ell}^{\pd}$, $c_{\ell}^{\pd}$, and $d_{\ell}^{\pd}$, and the vacuum state $| 0 \rangle = |\mathbf{0},\mathbf{0},\mathbf{0},\mathbf{0}\rangle$,
\begin{align}
    a_{\ell} | 0 \rangle = b_{\ell} | 0 \rangle = c_{\ell} | 0 \rangle = d_{\ell} | 0 \rangle = 0. 
\end{align}
Meanwhile, excited states $|\mathbf{m},\mathbf{n},\mathbf{o},\mathbf{p}\rangle$ are formed by the successive action of the relevant bosonic creation operators, with the standard definitions,
\begin{align}
    |\mathbf{m},\mathbf{n},\mathbf{o},\mathbf{p}\rangle = \: \prod_{\ell} & \frac{1}{\sqrt{m_{\ell}!}}\frac{1}{\sqrt{n_{\ell}!}}\frac{1}{\sqrt{o_{\ell}!}}\frac{1}{\sqrt{p_{\ell}!}}\times \nonumber \\
    & \left(a_{\ell}^{\dag}\right)^{m_{\ell}}\left(b_{\ell}^{\dag}\right)^{n_{\ell}}\left(c_{\ell}^{\dag}\right)^{o_{\ell}}\left(d_{\ell}^{\dag}\right)^{p_{\ell}}|0\rangle
\end{align}
with the action of the creation and annihilation operators defined as 
\begin{align}
        a_{\ell}^{\dag} |\mathbf{m},\mathbf{n},\mathbf{o},\mathbf{p}\rangle = \: & \sqrt{m_{\ell} + 1}|\mathbf{m}_{\ell}^{+},\mathbf{n},\mathbf{o},\mathbf{p}\rangle \\
        a_{\ell}^{\pd} |\mathbf{m},\mathbf{n},\mathbf{o},\mathbf{p}\rangle = \: & \sqrt{m_{\ell}} |\mathbf{m}_{\ell}^{-},\mathbf{n},\mathbf{o},\mathbf{p}\rangle,
\end{align}
where $\mathbf{m}_{\ell}^{\pm} = (m_{1},\dots,m_{\ell}\pm1,\dots)$. Analogous relations are also defined for $b_{\ell}^{\dag}$, $c_{\ell}^{\dag}$, and $d_{\ell}^{\dag}$. These definitions imply orthonormality between the auxiliary Fock states: 
\begin{align}
    \langle \mathbf{m}',\mathbf{n}',\mathbf{o}',\mathbf{p}'|\mathbf{m},\mathbf{n},\mathbf{o},\mathbf{p}\rangle = \: & \delta_{\mathbf{m}\mathbf{m}'}\delta_{\mathbf{n}\mathbf{n}'}\delta_{\mathbf{o}\mathbf{o}'}\delta_{\mathbf{p}\mathbf{p}'}
\end{align}

However, this is not the only representation available. The definition of the auxiliary Fock space also implies the existence of a corresponding coordinate representation, obtained by defining generalized canonical coordinate and momentum operators,
\begin{align}
	a_{\ell}&=\hat{\partial}_{q_{\ell}}+\frac{\hat{q}_{\ell}}{2} \:\:\:\: ; \:\:\:\: a_{\ell}^\dagger=-\hat{\partial}_{q_{\ell}}+\frac{\hat{q}_{\ell}}{2} \nonumber\\
	b_{\ell}&=\hat{\partial}_{p_{\ell}}+\frac{\hat{p}_{\ell}}{2} \:\:\:\: ; \:\:\:\: b_{\ell}^\dagger=-\hat{\partial}_{p_{\ell}}+\frac{\hat{p}_{\ell}}{2} \nonumber\\
	c_{\ell}&=\hat{\partial}_{r_{\ell}}+\frac{\hat{r}_{\ell}}{2} \:\:\:\: ; \:\:\:\: c_{\ell}^\dagger=-\hat{\partial}_{r_{\ell}}+\frac{\hat{r}_{\ell}}{2} \nonumber\\
	d_{\ell}&=\hat{\partial}_{s_{\ell}}+\frac{\hat{s}_{\ell}}{2} \:\:\:\: ; \:\:\:\: d_{\ell}^\dagger=-\hat{\partial}_{s_{\ell}}+\frac{\hat{s}_{\ell}}{2}, \label{eq: Fock to coordinate representation}
\end{align}
with identity operator
\begin{align}
    \mathbb{I}_{\text{AF}} = \: & \sum_{\substack{\mathbf{m},\mathbf{n}\\\mathbf{o},\mathbf{p}}} | \mathbf{m},\mathbf{n},\mathbf{o},\mathbf{p} \rangle \langle \mathbf{m},\mathbf{n},\mathbf{o},\mathbf{p} | \\
    = \: & \int d\mathbf{q}d\mathbf{p}d\mathbf{r}d\mathbf{s} \: | \mathbf{q},\mathbf{p},\mathbf{r},\mathbf{s} \rangle \langle \mathbf{q},\mathbf{p},\mathbf{r},\mathbf{s} |.
\end{align}
This choice of coordinate and momentum operators imposes the commutation relations
\begin{align}
    \left[a^{\pd}_{\ell},a^{\dag}_{\ell}\right] = \left[b^{\pd}_{\ell},b^{\dag}_{\ell}\right] = \left[c^{\pd}_{\ell},c^{\dag}_{\ell}\right] = \left[d^{\pd}_{\ell},d^{\dag}_{\ell}\right] = 1.
\end{align}
We are particularly interested in a representation in the generalized coordinate eigenbasis, $| \mathbf{q},\mathbf{p},\mathbf{r},\mathbf{s} \rangle $, formed from eigenstates of the position operators, with, for example $\mathbf{q} = (q_{1},\dots,q_{\ell},\dots,q_{N^{\text{max}}_{\ell}})$ and 
\begin{align}
    \hat{q}_{\ell} | q_{\ell} \rangle = \: & q_{\ell} | q_{\ell} \rangle, \label{eq: commutator relation auxiliary Fock space}
\end{align}
and equivalent relations for $\mathbf{p}$, $\mathbf{r} $, and $\mathbf{s}$.

The overlap between the coordinate eigenstates and the corresponding Fock states is given by the probabilist's Hermite polynomials,
\begin{align}
    & \langle \mathbf{q},\mathbf{p},\mathbf{r},\mathbf{s}  | \mathbf{m},\mathbf{n},\mathbf{o},\mathbf{p}  \rangle \nonumber \\ 
    & \: = \: \frac{1}{(2\pi)^{N^{\text{max}}_{\ell}}}\prod_{\ell}\frac{\text{He}_{m_{\ell}}(q_{\ell})\text{He}_{n_{\ell}}(p_{\ell})\text{He}_{o_{\ell}}(r_{\ell})\text{He}_{p_{\ell}}(s_{\ell})}{\sqrt{m_{\ell}!n_{\ell}!o_{\ell}!p_{\ell}!}} \times \nonumber \\
    & \qquad e^{-S(\mathbf{q},\mathbf{p},\mathbf{r},\mathbf{s})}, \label{eq: overlap identity}
\end{align}
where 
\begin{align}
    e^{-S(\mathbf{q},\mathbf{p},\mathbf{r},\mathbf{s})} = \: & \langle \mathbf{q},\mathbf{p},\mathbf{r},\mathbf{s} |e^{-S}| \mathbf{q},\mathbf{p},\mathbf{r},\mathbf{s} \rangle \nonumber \\
    = \: &e^{-\frac{|\mathbf{q}|^{2}+|\mathbf{p}|^{2}+|\mathbf{r}|^{2}+|\mathbf{s}|^{2}}{4}}
\end{align} 
is the Gaussian-weighting operator in the coordinate basis. Consequently, in this basis, Fock states are given by a Gaussian integral over the coordinate representation, with the vacuum state, 
\begin{align}
    |0\rangle = \frac{1}{(2\pi)^{N^{\text{max}}_{\ell}}} \int d\mathbf{q}d\mathbf{p}d\mathbf{r}d\mathbf{s} \: e^{-S(\mathbf{q},\mathbf{p},\mathbf{r},\mathbf{s})} | \mathbf{q},\mathbf{p},\mathbf{r},\mathbf{s} \rangle ,
\end{align}
and excited states,
\begin{align}
    |\mathbf{m},\mathbf{n},\mathbf{o},\mathbf{p}\rangle = \: & \frac{1}{(2\pi)^{N^{\text{max}}_{\ell}}} \int d\mathbf{q}d\mathbf{p}d\mathbf{r}d\mathbf{s} \:  | \mathbf{q},\mathbf{p},\mathbf{r},\mathbf{s} \rangle \prod_{\ell=1}^{N^{\text{max}}_{\ell}} \times \nonumber \\
    & \frac{\text{He}_{m_{\ell}}(q_{\ell})\text{He}_{n_{\ell}}(p_{\ell})\text{He}_{o_{\ell}}(r_{\ell})\text{He}_{p_{\ell}}(s_{\ell})}{\sqrt{2^{M_{\ell}+N_{\ell}+O_{\ell}+P_{\ell}}}\sqrt{m_{\ell}!n_{\ell}!o_{\ell}!p_{\ell}!}} \times \\
    & e^{-S(\mathbf{q},\mathbf{p},\mathbf{r},\mathbf{s})} . 
\end{align}

Now, the position-space representation of the extended density operator is given by
\begin{align}
    \rho(\mathbf{q},\mathbf{p},\mathbf{r},\mathbf{s}) = \: & \langle \mathbf{q},\mathbf{p},\mathbf{r},\mathbf{s} | \rho_{\text{ext}} \rangle\rangle \\
    = \: & \sum_{\substack{\mathbf{m}|\mathbf{n}\\ \mathbf{o}|\mathbf{p}}} \langle \mathbf{q},\mathbf{p},\mathbf{r},\mathbf{s} |\mathbf{m},\mathbf{n},\mathbf{o},\mathbf{p}\rangle\langle \mathbf{m},\mathbf{n},\mathbf{o},\mathbf{p}| \rho_{\text{ext}} \rangle \rangle \\
    = \: & \frac{1}{(2\pi)^{N^{\text{max}}_{\ell}}}\sum_{\substack{\mathbf{m}|\mathbf{n}\\ \mathbf{o}|\mathbf{p}}} \prod_{\ell} \times \nonumber \\
    & \frac{\text{He}_{m_{\ell}}(q_{\ell})\text{He}_{n_{\ell}}(p_{\ell})\text{He}_{o_{\ell}}(r_{\ell})\text{He}_{p_{\ell}}(s_{\ell})}{\sqrt{m_{\ell}!}\sqrt{n_{\ell}!}\sqrt{o_{\ell}!}\sqrt{p_{\ell}!}}e^{-S}\rho_{\substack{\mathbf{m}|\mathbf{n}\\ \mathbf{o}|\mathbf{p}}}.
\end{align}


\section{Stability Analysis of Simplified Model} \label{app: Stability Analysis of Simplified Model}

We aim to derive an estimate of the transient $2$-norm of the truncated propagator for the simplified model outlined in Eq.\eqref{eq: reduced generator}. By employing the Dunford-Cauchy representation of the propagator, 
\begin{align}
	e^{tA} = \: & \frac{1}{2\pi i} \oint_\Gamma e^{tz} (z - A)^{-1} dz,
\end{align}
the transient amplification can be related to the norm of the resolvent $\|(zI-A)^{-1}\|$. In the simplified model, the relevant non-normal direction is the auxiliary occupation chain generated by repeated application of $a^\dagger$, namely
\begin{align}
	|0\rangle \rightarrow |1\rangle \rightarrow \cdots \rightarrow |N\rangle .
\end{align}
This structure appears explicitly in the resolvent matrix element connecting the lowest and highest retained occupations,
\begin{align}
	\left[(zI-L_N)^{-1}\right]_{N,0} = \: & \frac{\eta^N\sqrt{N!}}{\prod_{k=0}^{N}(z+k\gamma)},
\end{align}
up to the convention used to order the occupation basis. 
Consequently,
  \begin{align}
	\|(zI-L_N)^{-1}\| \ge \: & \frac{|\eta|^N\sqrt{N!}}{\prod_{k=0}^{N}|z+k\gamma|}.	\label{eq: toy resolvent bound}
\end{align} 

For $z>0$ and $z\ll\gamma$, Eq.~\eqref{eq: toy resolvent bound} gives
\begin{align}
	\|(zI-L_N)^{-1}\| \gtrsim \: & \frac{1}{z}	\frac{(|\eta|/\gamma)^N}{\sqrt{N!}}.	\label{eq: toy resolvent approximation}
\end{align} 
For a normal operator with the same spectrum,
\begin{align}
	\|(zI-L_N)^{-1}\| = \: & \frac{1}{\mathrm{dist}\!\left(z,\sigma(L_N)\right)} \\
	= \: & \frac{1}{z},
\end{align}
such that the additional factor $(|\eta|/\gamma)^N/\sqrt{N!}$ directly reflects the resolvent amplification associated with the non-normal auxiliary occupation chain. For sufficiently large $|\eta|/\gamma$, this factor can grow rapidly with hierarchy depth over the relevant range of truncation tiers, showing that increasing the size of the finite auxiliary space does not necessarily suppress the sensitivity of the truncated generator.
  
The resolvent bound can also be related directly to the amplification of the propagator. For $z>0$,
\begin{align}
	(zI-L_N)^{-1} = \: & \int_0^\infty dt\, e^{-zt}e^{tL_N},
\end{align}
and therefore
\begin{align}
	\|(zI-L_N)^{-1}\| \le \: & \int_0^\infty dt\,e^{-zt} \sup_{t'\geq0}\|e^{t'L_N}\| = \frac{1}{z}	\sup_{t'\geq0}\|e^{t'L_N}\|.
\end{align}
Combining this relation with Eq.~\eqref{eq: toy resolvent approximation} yields 
\begin{align}
	\sup_{t\geq0}\|e^{tL_N}\| \gtrsim \: & \frac{(|\eta|/\gamma)^N}{\sqrt{N!}}.	\label{eq: structural insight}
\end{align}

\section{Perspective on Instabilities in the Coordinate Representation} \label{app: Perspective on Instability in the Coordinate Representation}

Although this simple Fock-space picture offers deep insight into the structural instabilities of bosonic HEOM, these issues can also be observed in other representations \cite{Ikeda2020_generalization,Li2022_a_low-temperature}. Specifically, as shown in Appendix~\ref{app: Details of the Auxiliary Fock Space}, the auxiliary Fock representation implies the existence of a corresponding extended coordinate representation $|\mathbf q,\mathbf p,\mathbf r,\mathbf s \rangle$, where the generalized coordinate and momentum operators in the auxiliary space are related to the generalized creation and annihilation operators explicitly in  Eq.\eqref{eq: Fock to coordinate representation}. In this representation, the extended state is given by 
\begin{align}
    \rho(\mathbf{q},\mathbf{p},\mathbf{r},\mathbf{s}) = \: & \langle\langle \mathbf{q},\mathbf{p},\mathbf{r},\mathbf{s} | \rho_{\text{ext}} \rangle\rangle. \label{eq: coordinate basis extended density operator}
\end{align}
By projecting $\mathcal{L}$ onto this basis and writing the annihilation and creation operators in the auxiliary Fock space with their corresponding generalized coordinate representations, one would obtain a Fokker-Planck equation of motion for $\rho(\mathbf{q},\mathbf{p},\mathbf{r},\mathbf{s})$. In this formulation, the generator $\mathcal{L}$ takes the form of a Fokker-Planck-type operator consisting of drift, diffusion, and mixed derivative terms in the hierarchy coordinates. 


From this perspective, the same instability mechanism can be represented in terms of poorly conditioned drift-diffusion dynamics in the auxiliary coordinate space. The resulting pseudospectral growth in the truncated hierarchy can thus be interpreted as the coordinate-space analogue of non-normal amplification, arising from the interplay between drift-dominated transport and the artificial boundary conditions introduced by truncation. This suggests a natural strategy for mitigating these instabilities: reformulating the dynamics in a representation in which drift and diffusion contributions are more evenly balanced, thereby reducing the mechanisms responsible for transient non-normal growth.

\section{Coordinate Representation of Bogoliubov-Transformed Operators} \label{app: Coordinate Representation of Bogoliubov-Transformed Operators}

The annihilation and creation operators defined in Eq.\eqref{eq: Bog operators} can be written in the same coordinate basis as the original Fock operators, albeit with different definitions:
\begin{align}
	A_{\ell}& = \frac{1}{\sqrt{2}}\left(\partial_{q_{\ell}}+q_{\ell} \right) \qquad A_{\ell}^\dagger =\frac{1}{\sqrt{2}}\left(-\partial_{q_{\ell}}+q_{\ell} \right) \nonumber	\\
	B_{\ell}&=\frac{1}{\sqrt{2}}\left(\partial_{p_{\ell}}+p_{\ell} \right) \qquad B_{\ell}^\dagger=\frac{1}{\sqrt{2}}\left(-\partial_{p_{\ell}}+p_{\ell} \right)  \nonumber\\
	C_{\ell}&=\frac{1}{\sqrt{2}}\left(\partial_{r_{\ell}}+r_{\ell} \right) \qquad C_{\ell}^\dagger=\frac{1}{\sqrt{2}}\left(-\partial_{r_{\ell}}+r_{\ell} \right)  \nonumber\\
	D_{\ell}&=\frac{1}{\sqrt{2}}\left(\partial_{s_{\ell}}+s_{\ell} \right) \qquad D_{\ell}^\dagger=\frac{1}{\sqrt{2}}\left(-\partial_{s_{\ell}}+s_{\ell} \right).
\end{align}
We denote the Fock-state basis of $\{A_{\ell},B_{\ell},C_{\ell},D_{\ell}\}$ as $\{|\mathbf{M},\mathbf{N},\mathbf{O},\mathbf{P}\rangle\}$, with the vacuum state labeled $|0\rangle_{\text{Bog.}} = |\mathbf{0},\mathbf{0},\mathbf{0},\mathbf{0}\rangle_{\text{Bog.}}$ to avoid confusion with the vacuum state of the untransformed operators. As before, it is defined as 
\begin{align}
    A_{\ell}|0\rangle_{\text{Bog.}} = B_{\ell}|0\rangle_{\text{Bog.}} = C_{\ell}|0\rangle_{\text{Bog.}} = D_{\ell}|0\rangle_{\text{Bog.}} = 0. 
\end{align}
We obtain the vacuum state in the coordinate representation, $\psi_{0,\text{Bog.}}( \mathbf{q},\mathbf{p},\mathbf{r},\mathbf{s})$ by inserting $\mathbb{I}_{\text{AF}}$, considering just a single $A_{\ell}$ explicitly,
\begin{align}
    A_{\ell}|0\rangle_{\text{Bog.}} = \: & \int d\mathbf{q}d\mathbf{p}d\mathbf{r}d\mathbf{s} \:  A_{\ell}| \mathbf{q},\mathbf{p},\mathbf{r},\mathbf{s} \rangle \langle \mathbf{q},\mathbf{p},\mathbf{r},\mathbf{s} | 0 \rangle \\
    = \: & \int dq_{\ell} \: | \mathbf{q},\mathbf{p},\mathbf{r},\mathbf{s} \rangle  \left(\partial_{q_{\ell}}+q_{\ell}\right) \mathsf{M}_{0,\text{Bog.}}( \mathbf{q},\mathbf{p},\mathbf{r},\mathbf{s}) \\
    \overset{!}{=} \: & 0.
\end{align}
Repeating this process for all $\{A_{\ell},B_{\ell},C_{\ell},D_{\ell}\}$, one obtains for the transformed vacuum state in the coordinate representation
\begin{align}
    \psi_{0,\text{Bog.}}( \mathbf{q},\mathbf{p},\mathbf{r},\mathbf{s}) = \: & \frac{1}{\left(\pi\right)^{N^{\text{max}}_{\ell}}} e^{-2S(\mathbf{q},\mathbf{p},\mathbf{r},\mathbf{s})} 
\end{align}
where the factor of $\left(\pi\right)^{-N^{\text{max}}_{\ell}}$ is necessary for normalization. 

The excited states $\{|\mathbf{M},\mathbf{N},\mathbf{O},\mathbf{P}\rangle\}$ satisfy the standard Fock representation,
\begin{align}
    |\mathbf{M},\mathbf{N},\mathbf{O},\mathbf{P}\rangle = \: \prod_{\ell} & \frac{1}{\sqrt{M_{\ell}!}}\frac{1}{\sqrt{N_{\ell}!}}\frac{1}{\sqrt{O_{\ell}!}}\frac{1}{\sqrt{P_{\ell}!}}\times \nonumber \\
    & \left(A_{\ell}^{\dag}\right)^{M_{\ell}}\left(B_{\ell}^{\dag}\right)^{N_{\ell}}\left(C_{\ell}^{\dag}\right)^{O_{\ell}}\left(D_{\ell}^{\dag}\right)^{P_{\ell}}|0\rangle_{\text{Bog.}}
\end{align}
with the action of the creation and annihilation operators given by 
\begin{align}
        A_{\ell}^{\dag} |\mathbf{M},\mathbf{N},\mathbf{O},\mathbf{P}\rangle = \: & \sqrt{M_{\ell} + 1}|\mathbf{M}_{\ell}^{+},\mathbf{N},\mathbf{O},\mathbf{P}\rangle \\
        A_{\ell}^{\pd} |\mathbf{M},\mathbf{N},\mathbf{O},\mathbf{P}\rangle = \: & \sqrt{M_{\ell}} |\mathbf{M}_{\ell}^{-},\mathbf{N},\mathbf{O},\mathbf{P}\rangle,
\end{align}
where $\mathbf{M}_{\ell}^{\pm} = (M_{1},\dots,M_{\ell}\pm 1,\dots)$. Analogous relations are also defined for $B_{\ell}^{\dag}$, $C_{\ell}^{\dag}$, and $D_{\ell}^{\dag}$. These definitions imply orthonormality between the Fock states: 
\begin{align}
    \langle \mathbf{M}',\mathbf{N}',\mathbf{O}',\mathbf{P}'|\mathbf{M},\mathbf{N},\mathbf{O},\mathbf{P}\rangle = \: & \delta_{\mathbf{M}\mathbf{M}'}\delta_{\mathbf{N}\mathbf{N}'}\delta_{\mathbf{O}\mathbf{O}'}\delta_{\mathbf{P}\mathbf{P}'}
\end{align}

However, in the coordinate representation, it can be shown that the Bogoliubov-transformed excited states must be expressed in terms of the physicist's Hermite polynomials, $\text{H}_{M_{\ell}}(q_{\ell})$, rather than the probabilist's Hermite polynomials,
\begin{align}
    |\mathbf{M},\mathbf{N},\mathbf{O},\mathbf{P}\rangle = \: & \frac{1}{ \pi^{N^{\text{max}}_{\ell}}} \int d\mathbf{q}d\mathbf{p}d\mathbf{r}d\mathbf{s} \: | \mathbf{q},\mathbf{p},\mathbf{r},\mathbf{s} \rangle \prod_{\ell} \times \nonumber \\
    & \frac{\text{H}_{M_{\ell}}(q_{\ell})\text{H}_{N_{\ell}}(p_{\ell})\text{H}_{O_{\ell}}(r_{\ell})\text{H}_{P_{\ell}}(s_{\ell})}{\sqrt{2^{M_{\ell}+N_{\ell}+O_{\ell}+P_{\ell}}}\sqrt{M_{\ell}!N_{\ell}!O_{\ell}!P_{\ell}!}} \times \nonumber \\
    & e^{-2S(\mathbf{q},\mathbf{p},\mathbf{r},\mathbf{s})} . \label{eq: Bog. transformed coordinate representation}
\end{align}

Finally, we note that, in terms of these rotated ladder operators, the transformed HEOM generator takes the compact form 
  \begin{align}
    \mathcal{L} = \: & -i\mathcal{L}_{\text{S}} - \sum_{\ell = 1}^{N^{\text{max}}_{\ell}} \gamma_{\ell,r} \left[ \sum_{\alpha \in\{A,B,C,D\}}\left(\alpha^{\dag}_{\ell}\alpha^{\pd}_{\ell} - \alpha^{\pd}_{\ell}\alpha^{\pd}_{\ell}\right)\right] - \nonumber \\
    & \sqrt{2}i \sum_{\ell = 1}^{N^{\text{max}}_{\ell}} \mathcal{L}_{O_{S}} \sum_{\alpha \in\{A,B,C,D\}} \alpha^{\pd}_{\ell} \: - \nonumber \\
    & \frac{i}{\sqrt{2}} \sum_{\ell = 1}^{N^{\text{max}}_{\ell}} \left(\eta^{\pd}_{\ell}O_{S}^{L} \left(A^{\dag}_{\ell} - A^{\pd}_{\ell}\right) - \eta^{*}_{\ell} O_{S}^{R} \left(C^{\dag}_{\ell} - C^{\pd}_{\ell}\right)\right) - \nonumber \\
    & i \sum_{\ell = 1}^{N^{\text{max}}_{\ell}} \gamma_{\ell,i} \left(A^{\dag}_{\ell}B^{\pd}_{\ell} - 2A^{\pd}_{\ell}B^{\pd}_{\ell} + A^{\pd}_{\ell}B^{\dag}_{\ell}\right) +  \nonumber \\
    & i \sum_{\ell = 1}^{N^{\text{max}}_{\ell}} \gamma_{\ell,i} \left(C^{\dag}_{\ell}D^{\pd}_{\ell} - 2C^{\pd}_{\ell}D^{\pd}_{\ell} + C^{\pd}_{\ell}D^{\dag}_{\ell}\right). \label{eq: HEOM extended space transformed 2}
\end{align}

\section{Inverse Similarity Transformation} \label{app: Similarity Transformation}

The task is to evaluate the coefficients defining the inverse similarity transformation to obtain the reduced system density matrix in Eq.\eqref{eq: inverse similarity transformation rho_S},
\begin{align}
    \tilde{\Theta}^{-1}_{\substack{\mathbf{0}|\mathbf{0}\\ \mathbf{0}|\mathbf{0}};\substack{\mathbf{M}|\mathbf{N}\\ \mathbf{O}|\mathbf{P}}}  = \: &  \mathcal{N}_{0}^{-1}\left(\prod_{\ell} \frac{1}{\sqrt{M_{\ell}!\: N_{\ell}!\:O_{\ell}!\:P_{\ell}!}}\right) \times \nonumber \\
    & \langle \mathbf{0},\mathbf{0},\mathbf{0},\mathbf{0} | e^{+S} | \mathbf{M},\mathbf{N},\mathbf{O},\mathbf{P} \rangle.
\end{align}
This is easiest to evaluate in the coordinate representation of the extended space. Inserting the representation of the Bogoliubov-transformed Fock states in the coordinate representation in Eq.\eqref{eq: Bog. transformed coordinate representation} yields
\begin{align}
    \tilde{\Theta}^{-1}_{\substack{\mathbf{0}|\mathbf{0}\\ \mathbf{0}|\mathbf{0}};\substack{\mathbf{M}|\mathbf{N}\\ \mathbf{O}|\mathbf{P}}}  = \: & \frac{1}{(4\pi^{2})^{N^{\text{max}}_{\ell}}} \int d\mathbf{q}d\mathbf{p}d\mathbf{r}d\mathbf{s} \: e^{+S(\mathbf{q},\mathbf{p},\mathbf{r},\mathbf{s})} \prod_{\ell} \times \nonumber \\
    & \frac{\text{H}_{M_{\ell}}(q_{\ell})\text{H}_{N_{\ell}}(p_{\ell})\text{H}_{O_{\ell}}(r_{\ell})\text{H}_{P_{\ell}}(s_{\ell})}{\sqrt{2^{M_{\ell}+N_{\ell}+O_{\ell}+P_{\ell}}}M_{\ell}!N_{\ell}!O_{\ell}!P_{\ell}!} e^{-3S(\mathbf{q},\mathbf{p},\mathbf{r},\mathbf{s})}.
\end{align}
Using the identity
\begin{align}
	\int_{-\infty}^{\infty} e^{-\frac{x^{2}}{2}} \text{H}_n (x) dx = 
	\begin{cases}
		\sqrt{2\pi} \: 2^{n/2}(n-1)!! \qquad & n \text{ even} \\
		0 \qquad & n \text{ odd}
	\end{cases},
\end{align}
one obtains 
\begin{widetext}
	\begin{align}
		\tilde{\Theta}^{-1}_{
			\substack{\mathbf{0}|\mathbf{0}\\
				\mathbf{0}|\mathbf{0}};
			\substack{\mathbf{M}|\mathbf{N}\\
				\mathbf{O}|\mathbf{P}}
		}
		=
		\left\{
		\begin{array}{ll}
			\displaystyle
			\prod\limits_{\ell}
			\frac{
				(M_{\ell}-1)!!
				(N_{\ell}-1)!!
				(O_{\ell}-1)!!
				(P_{\ell}-1)!!
			}{
				M_{\ell}!\,
				N_{\ell}!\,
				O_{\ell}!\,
				P_{\ell}!
			},
			&
			M_{\ell},N_{\ell},O_{\ell},P_{\ell}
			\text{ all even},
			\\[3mm]
			0,
			&
			\text{otherwise}.
		\end{array}
		\right.
	\end{align}
\end{widetext}

\clearpage

\bibliography{Main_text_incl._fig.bib}

\end{document}